\documentclass[sigconf,10pt,nonacm]{acmart}

\usepackage{enumitem}
\usepackage{xspace}
\newcommand{\sysname}{\texttt{AirLog}\xspace}

\usepackage{booktabs}
\usepackage{multirow}
\usepackage{calc}
\newlength{\topheight}
\usepackage{listings}
\usepackage[table]{xcolor}
\usepackage{subcaption}
\usepackage{placeins}
\usepackage{algorithmic}
\usepackage{graphicx}
\usepackage{textcomp}
\usepackage{xcolor}
\hypersetup{colorlinks=true}

\AtBeginDocument{%
  }

\begin{document}

\title{\sysname: Store-Level Indoor Life Logging Made Easy}

\author{Zihui Yun}
\affiliation{%
  \institution{University of Georgia}
  \city{Athens}
  \state{Georgia}
  \country{USA}}
\email{ayucharon666@gmail.com}

\author{Jiaying Du}
\affiliation{%
  \institution{University of Georgia}
  \city{Athens}
  \state{Georgia}
  \country{USA}}
\email{msz@uga.edu}

\author{Yue Yu}
\affiliation{%
  \institution{University College London}
  \city{London}
  \country{United Kingdom}}
\email{ucfnyul@ucl.ac.uk}

\author{Zhewei Liu}
\affiliation{%
  \institution{University of Toronto Mississauga}
  \city{Mississauga}
  \state{Ontario}
  \country{Canada}}
\email{zwei.liu@utoronto.ca}

\author{Zhen Xiang}
\affiliation{%
  \institution{University of Georgia}
  \city{Athens}
  \state{Georgia}
  \country{USA}}
\email{zxiangaa@uga.edu}

\author{Longfei Shangguan}
\affiliation{%
  \institution{University of Pittsburgh}
  \city{Pittsburgh}
  \state{Pennsylvania}
  \country{USA}}
\email{longfei@pitt.edu}

\author{Zhenlin An}
\authornote{Corresponding author.}
\affiliation{%
  \institution{University of Georgia}
  \city{Athens}
  \state{Georgia}
  \country{USA}}
\email{zhenlin.an@uga.edu}

\renewcommand{\shortauthors}{Yun et al.}

\begin{abstract}
This paper presents \sysname, a smartphone-based life journaling system that automatically reconstructs users' store visits in shopping malls and summarizes them into human-readable journals.
Unlike conventional indoor localization systems, \sysname avoids labor-intensive radio-map construction and dedicated wireless localization infrastructure and algorithm calibrations. Instead, it repurposes two cues already available in commercial spaces: semantic information exposed by ambient Wi-Fi SSIDs and indoor directory images. \sysname converts directory images into spatial maps and fuses Wi-Fi semantic anchors with inertial dead reckoning to recover store-level trajectories, which are then summarized into journals by an LLM.
Such store-level life logs can support applications such as personal memory recall, activity reflection, and automated diary generation without requiring users to manually record where they have been.
We implement \sysname on commodity smartphones and evaluate it on both a
large-scale public dataset and a self-collected dataset. The results demonstrate that \sysname substantially improves store-level region recovery, semantic matching, trajectory reconstruction, and journal quality over existing baselines. A human evaluation further shows that the generated journals are coherent and faithful to users' visits.
\end{abstract}
\maketitle

\section{Introduction}

\begin{figure}[t!]
    \centering
    \includegraphics[width=\linewidth]{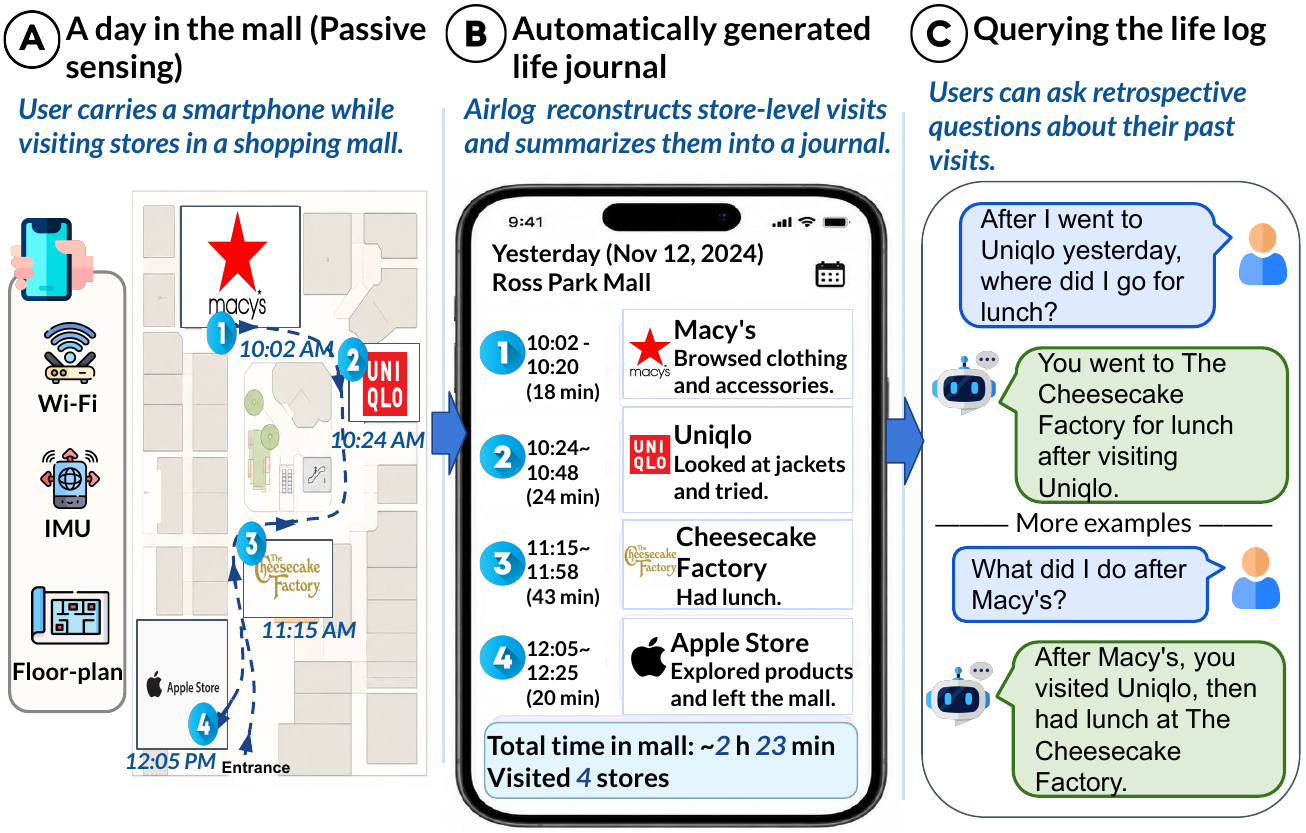}
    \vspace{-7mm}
    \caption{\textbf{\sysname turns passive smartphone sensing into a store-level indoor life log.}
\textnormal{(A) While the user visits a shopping mall, the smartphone passively collects Wi-Fi scans and inertial measurements, and a publicly available floor-plan image is used as spatial context. (B) \sysname reconstructs the store-level trajectory—what stores were visited, in what order, and for how long—and summarizes it into a human-readable journal. (C) The life log is queryable: users can ask retrospective questions, such as where they went for lunch after visiting a specific store.}}
    \label{fig:scenario}
    \vspace{-7mm}
\end{figure}

Automatically recording daily experiences has long been a goal of mobile sensing and context-aware computing~\cite{gurrin2014lifelogging,gemmell2006mylifebits}. In this paper, we focus on \emph{store-level indoor life logging}: automatically reconstructing which stores or venues a user visited in a shopping mall, in what order, and for how long, and summarizing these visits into a human-readable journal. Such logs can support personal memory recall, visit retrospection, and context-aware assistants. 
As illustrated in Figure~\ref{fig:scenario}, instead of merely
recording that a user spent two hours at a mall, a useful log could
report that the user visited Macy's, then Uniqlo, had lunch at the
Cheesecake Factory, and later stopped by the Apple Store, together with
the temporal order and dwell time of these visits. When users cannot
remember the details of a past visit, they can also directly query the
life log, e.g., ``Where did I go for lunch after visiting Uniqlo?''

At first glance, this task appears to be a straightforward extension of existing mobile lifelogging systems where coarse-grained mobility context is often  obtained directly from GPS traces, digital maps, and nearby points of interest (POIs), after which an LLM can interpret or summarize the resulting context~\cite{yu2025sensorchat,ouyang2024llmsense,xu2024penetrative,xu2025autolife,post2025contextllm}.
Indoors, however, the store-level trajectory itself is not readily available, largely because GPS is unreliable while conventional Wi-Fi localization typically requires site-specific surveys, fingerprint databases, or dedicated localization pipelines~\cite{bahl2000radar,youssef2005horus,ni2022experience,bhatia2025transforming,harle2013survey,zafari2019survey}. Such requirements make store-level life logging difficult to scale across large and frequently changing commercial spaces.

Our key observation is that commercial indoor environments already expose two useful spatial cues: (i): {\it ambient Wi-Fi SSIDs}, which often reveal semantic information about nearby stores~\cite{seneviratne2015ssids,korelivc2025sellma,zhang2025wichat}; and (ii): {\it indoor directory images}, which expose region locations and readable labels. 
This leads to our central question: \emph{Can these readily available but weak semantic cues be combined with smartphone motion sensing to recover store-level trajectories without relying on traditional Wi-Fi radio map or localization infrastructures?}

Answering this question introduces two challenges: 
\begin{enumerate}[leftmargin=*]
    \item \emph{public indoor maps are rarely available as machine-readable region layers}. While outdoor map services provide structured roads, venue boundaries, and POIs, indoor floor-level information often exists only as directory images rather than labeled region geometry that downstream modules can query~\cite{googleIndoorMapsHelp,googleIndoorMapsAbout,ogcIndoorGML,ogcIMDF2021}. Direct reasoning over these images is unreliable because current vision-language models (VLMs) still struggle with dense map text and fine-grained spatial relationships~\cite{xing2025mapbench}.
    \item \emph{Ambient Wi-Fi provides semantic but inherently ambiguous location evidence}. An SSID may reveal that a particular store is nearby, but it does not uniquely determine whether the user is inside that store, walking past it, or located in an adjacent venue. RSSI further varies with multipath, crowd blockage, device orientation, and heterogeneous AP placement. Thus, Wi-Fi should be treated as a probabilistic semantic anchor rather than a direct location estimate.
\end{enumerate}

\begin{figure}[t!]
    \centering
    \includegraphics[width=\linewidth]{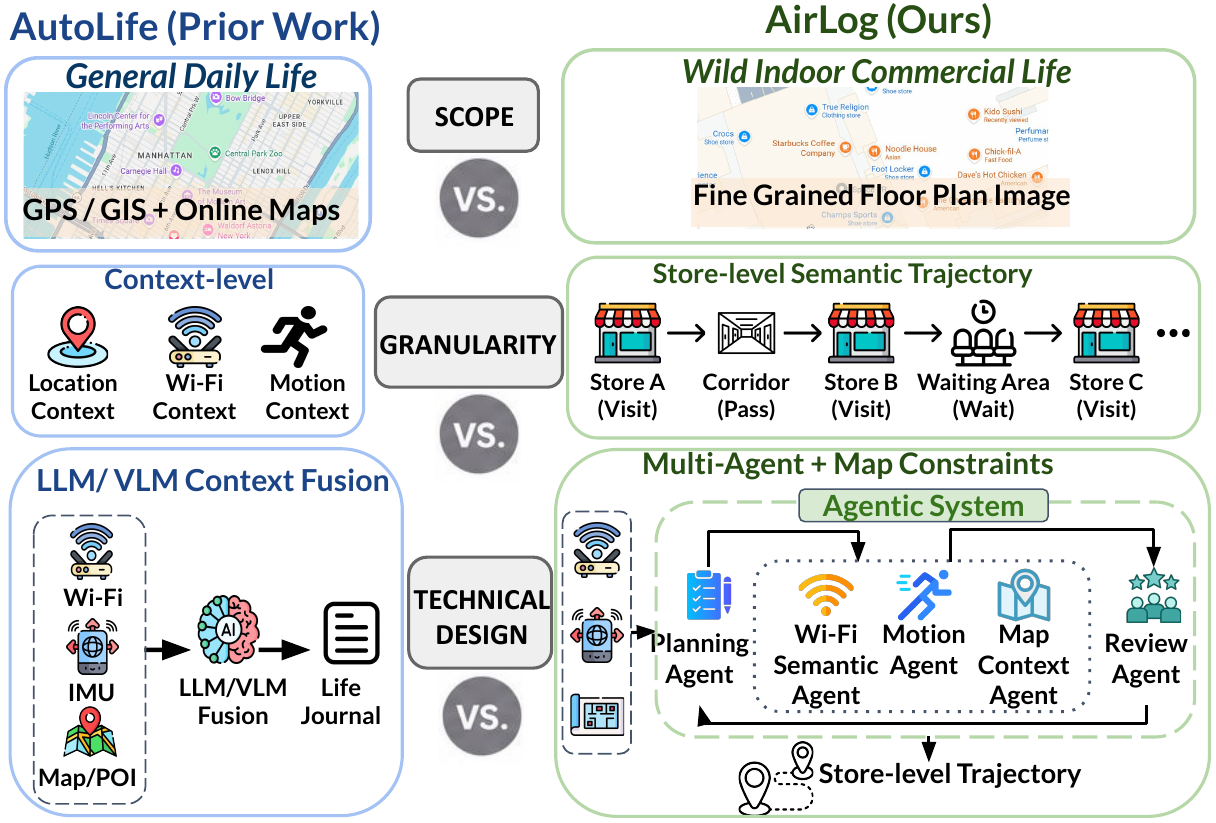}
    \vspace{-7mm}
    \caption{\textbf{From general daily life logging to store-level indoor life journaling.}
\textnormal{AutoLife captures coarse multimodal context, while \sysname uses structured indoor maps and map-constrained multi-agent reasoning to reconstruct store-level trajectories.}}
    \label{fig:comparison}
    \vspace{-6mm}
\end{figure}

\begin{figure}[t!]
    \includegraphics[width=\linewidth]{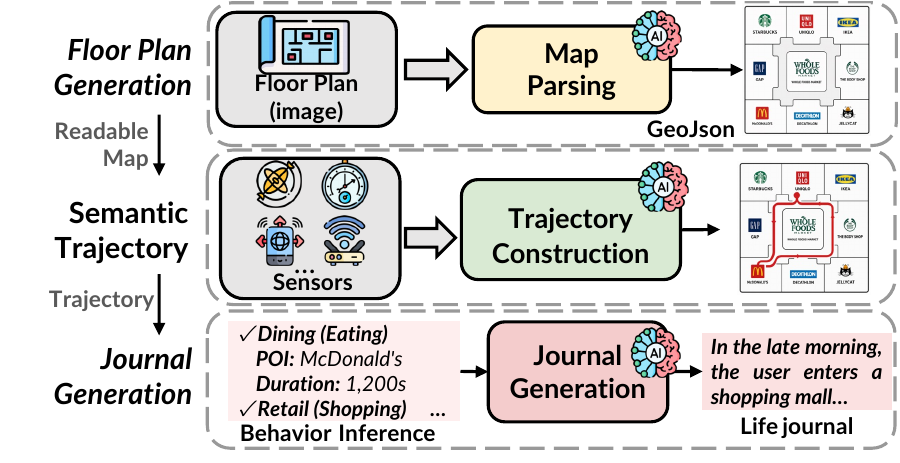}
    \vspace{-8mm}
\caption{\textbf{\sysname overview.} \textnormal{\sysname decomposes life logging into three structured layers: \textit{(1) Labeled Region Map Construction:} a geometry-first parser converts an indoor directory image into a labeled region map serialized as GeoJSON. \textit{(2) Trajectory Reconstruction:} a trajectory agent grounds sensor data and Wi-Fi semantics using downstream map constraints to infer store-level movement. \textit{(3) Journal Generation:} a journal agent summarizes the validated trajectory into a third-person life journal.}}
    \label{fig:overview}
    \vspace{-6mm}
\end{figure}

In this paper, we propose \sysname, a smartphone-based system for store-level indoor life journaling. As shown in Fig.~\ref{fig:overview}, given passive smartphone Wi-Fi scans, inertial motion recordings, and a crowd-accessible floor-plan image, \sysname infers which stores a user likely visited, in what order, and for how long, and then summarizes these visits into a life journal. Rather than asking an LLM to infer the entire journal from raw sensor traces and a floor-plan image, \sysname decomposes the task into three structured tasks.

\noindent $\bullet$ First, \emph{the map layer converts a human-facing indoor directory image into a labeled region map.} The parser serializes this map as a GeoJSON
\texttt{FeatureCollection}\footnote{GeoJSON is an industry-standard JSON-based
format for encoding map features and their geometry~\cite{rfc7946}. In
\sysname, each parser-produced feature represents one region using
\texttt{Polygon} or \texttt{MultiPolygon} geometry and stores its readable
strings in \texttt{properties.labels}.}, allowing later modules to
query region geometry and labels instead of reasoning over raw pixels. The
parser itself does not emit corridor polygons, walkable areas, or connectivity.
A separate deterministic map-preparation stage derives the corridor mask and
trajectory graph used by the fusion layer.

However, indoor directory images are designed for human viewing rather than
machine parsing. VLMs can miss dense labels or distort region boundaries, while
generic segmentation and traditional CV methods are brittle to text, icons,
reflections, and perspective distortion~\cite{carion2025sam3,su2022shoppingmallplans}.
\sysname therefore uses a geometry-first map parser: GPT Image 2 canonicalizes
the visual layout into a clean region mask, deterministic CV aligns and
vectorizes the regions, and GPT-4o labels one polygon-masked crop at a time
before writing the returned label set directly to the source GeoJSON feature.

\noindent $\bullet$ Second, \emph{the trajectory layer grounds Wi-Fi and motion observations onto the map to recover store visits.} 
Wi-Fi provides useful semantic hints: an SSID such as Uniqlo\_Guest may suggest a nearby Uniqlo. However, the SSID names can be ambiguous, abbreviated, and do not reveal whether the user actually entered the store, walked past it, or visited a neighboring store. 

\sysname therefore treats SSIDs as probabilistic semantic anchors rather than direct location estimates. It filters and aggregates Wi-Fi scans, derives pedestrian dead reckoning (PDR) motion from inertial measurements, and jointly evaluates candidate store sequences against map connectivity, walking distance, and motion continuity. The decoded semantic trajectory is then reviewed against these tool outputs; detected
semantic or physical inconsistencies are fed back into the loop to improve the final trajectory.

\noindent $\bullet$ Third, \emph{the journal layer converts the validated trajectory and motion sensor data into a human-readable account of the visit.} A store trajectory alone does not distinguish, for example, spending 25 minutes shopping inside a store from briefly passing its entrance. \sysname therefore combines dwell time, POI type, motion-derived activity cues, and trajectory confidence to derive a sequence of supported behavior events. An LLM then summarizes this validated sequence into a journal, rather than generating entries directly from raw Wi-Fi or sensor observations.

We evaluate \sysname with two main data sources: large-scale Microsoft
public indoor traces containing over 19.7k labeled Wi-Fi scan
samples~\cite{locationcompetition2020indoor}, and a self-collected six-city
corpus containing 374,567 Wi-Fi observations over 603.0 active hours. From the
latter, we construct 185,140 candidate SSID--POI pairs for semantic analysis.
The public traces support quantitative semantic trajectory recovery, while the
self-collected data support SSID semantic validation and end-to-end journal
generation validation.
We test \sysname's local deployment on edge devices, where sensing data can be collected during the day and processed locally afterward. On-device execution takes longer than cloud, but it preserves privacy by keeping raw data and journals on the user's device, while smaller models still achieve comparable accuracy.

\noindent In summary, this paper makes the following contributions:
\begingroup
\setlength{\topsep}{0pt}
\setlength{\partopsep}{0pt}
\setlength{\itemsep}{0.5pt}
\setlength{\parsep}{0pt}
\begin{itemize}[leftmargin=*,nosep]
    \item We propose \sysname, a novel agentic platform for survey-free,
    site services-free indoor life journaling in wild environments using
     smartphone sensing and crowd-accessible map artifacts.
    \item We design a geometry-first map parser that converts human-facing
    indoor directory images into labeled region maps serialized as GeoJSON, making
    region polygons and their readable labels directly queryable by downstream
    modules.
    \item We introduce a grounded semantic trajectory and journaling layer that
    treats ambient Wi-Fi SSIDs as semantic landmarks, fuses them with PDR and
    floor-plan topology through graph-constrained inference, and generates
    third-person journals from validated visits rather than raw sensor context.
\end{itemize}
\endgroup

\vspace{-5mm}
\section{Related Works}
\label{sec:related_works}

\subsection{Life Journaling}
Lifelogging systems transform daily observations into records for behavior understanding~\cite{gurrin2014lifelogging,gemmell2006mylifebits,bolanos2016toward,doherty2013passive}. Existing approaches broadly use either vision-based or non-visual sensing. Wearable cameras and smart glasses capture rich egocentric context~\cite{hodges2006sensecam,lee2008wearable,10611736,11095171,wang2025videoitg}, but incur high processing costs and privacy concerns, while smartphone-based approaches exploit motion, wireless, audio, and location signals~\cite{10.1145/3643553,kwapisz2011activity,shoaib2014fusion,yurur2014context}. Recent LLM-based systems further translate heterogeneous sensor data into open-vocabulary event descriptions~\cite{xu2024penetrative,10590466,ouyang2024llmsense,nepal2024mindscape,xu2025autolife,an2025iotllm,10697418,10.1145/3715014.3722082}, but largely rely on prompt-level context and target outdoor mobility or coarse activities. As illustrated in Fig.~\ref{fig:comparison}, AirLog instead grounds Wi-Fi and motion evidence on indoor map structure to recover store-level semantic trajectories before journal generation.

\vspace{-5mm}

\subsection{Indoor Localization}
Existing indoor localization methods~\cite{harle2013survey,zafari2019survey,alarifi2016ultra,decawave2017standard,qiao2024uwb} broadly follow fingerprinting- or model-based paradigms. Fingerprinting methods require site-specific radio measurements and machine learning~\cite{bahl2000radar,youssef2005horus,he2016wifi,bhatia2025transforming}, while ranging- or AoA-based methods depend on calibrated infrastructure and assumptions specific to each deployment~\cite{zafari2019survey,zhao2024util}, making both costly to deploy across changing commercial environments. Prior survey-free localization systems avoid manual site surveys through crowdsourcing, PDR, map constraints, and opportunistic landmarks, but still primarily target metric coordinates and typically benefit from observations accumulated across users or visits~\cite{rai2012zee,wang2012unloc,yang2012lifs,shen2013walkiemarkie}. \sysname instead targets \emph{site-free semantic localization}, inferring visited stores from a single visit without site-specific fingerprints or accumulated crowdsourced observations. Wi-Fi SSIDs provide a promising survey-free semantic signal by encoding business names and location-related tokens~\cite{seneviratne2015ssids,korelivc2025sellma,zhang2025wichat}; \sysname revalidates this signal at scale and fuses it with PDR and map topology to recover store-level trajectories rather than isolated place labels.

\subsection{Map Understanding with VLMs}
Recent work uses VLMs for map reading, landmark reasoning, and navigation from floor-plan images~\cite{zhou2023navgpt,li2025flona,chen2026floorplanvln}, yet current LVLMs struggle to jointly handle OCR, topology, and route planning~\cite{xing2025mapbench}. Meanwhile, public indoor-map services rarely expose floor-level geometry and topology as queryable layers~\cite{googleIndoorMapsHelp,googleIndoorMapsAbout,ogcIndoorGML,ogcIMDF2021}, and recovering queryable regions and labels from human-facing maps remains challenging. \sysname addresses these gaps by converting directory images into labeled GeoJSON region maps for reasoning over explicit polygons and labels. Existing map-parsing pipelines combine OCR and classical segmentation~\cite{liu2017raster2vec,yang2019floornet,su2022shoppingmallplans}, but remain brittle on photographed directories with text, icons, legends, reflections, and perspective distortion; \sysname instead canonicalizes region geometry with image generation, vectorizes polygons using deterministic CV, and labels each region with a per-region semantic subagent~\cite{visionbanana2026}.

\section{A Glimpse of \sysname}
\label{sec:AirLog}

Imagine Bob visiting Rose Park Mall on a Saturday afternoon. Before entering, he photographs the mall directory. He first enters \textit{Uniqlo} and browses, then walks past \textit{KFC}, and later spends about 25 minutes at \textit{Starbucks}. Bob never checks in or labels these visits; during the trip, his phone records only passive Wi-Fi scans and IMU readings.

To reconstruct Bob's path, \sysname must first convert this directory photo into a labeled region map (\S\ref{sec:floorplan}). Although readable to Bob, the raw image cannot directly serve as map context for the trajectory agents, which need explicit store geometry and POI labels. The resulting GeoJSON stores labeled region polygons for Uniqlo, KFC, and Starbucks so later modules can combine the map with Bob's Wi-Fi and motion observations.

As Bob walks, his phone periodically observes SSIDs, including \texttt{Uniqlo\_WiFi}, \texttt{KFC\_FREE\_WiFi}, and \texttt{Starbucks\_WiFi}. However, these SSID observations are sometimes ambiguous:
seeing \texttt{KFC\_FREE\_WiFi}, for example, does not imply that Bob entered
KFC: he may simply be walking past it. \sysname therefore treats Wi-Fi
observations as probabilistic semantic anchors, uses PDR to estimate how Bob
moves between observations, and jointly decodes them under the floor-plan
topology. Said differently, the Wi-Fi SSIDs indicate \emph{where Bob may be}, motion sensing
indicates \emph{how Bob moved}, and the map constrains \emph{where Bob can move}.
These cues together can help distinguish a brief pass near KFC from sustained visits
to Uniqlo and Starbucks (\S\ref{sec:spatial_fusion}).

Finally, \sysname aggregates the reconstructed trajectory into store-level
visits using dwell time and motion cues, and summarizes the validated sequence
into a human-readable journal, e.g.,
\textit{``In the afternoon, Bob visited Uniqlo, walked through the dining area,
and later spent about 25 minutes at Starbucks.''} (\S\ref{sec:diary_generation}).

\section{Constructing Labeled Region Maps from Floor-Plan Images}
\label{sec:floorplan}

\begin{figure}[t!]
    \centering
    \includegraphics[width=\linewidth]{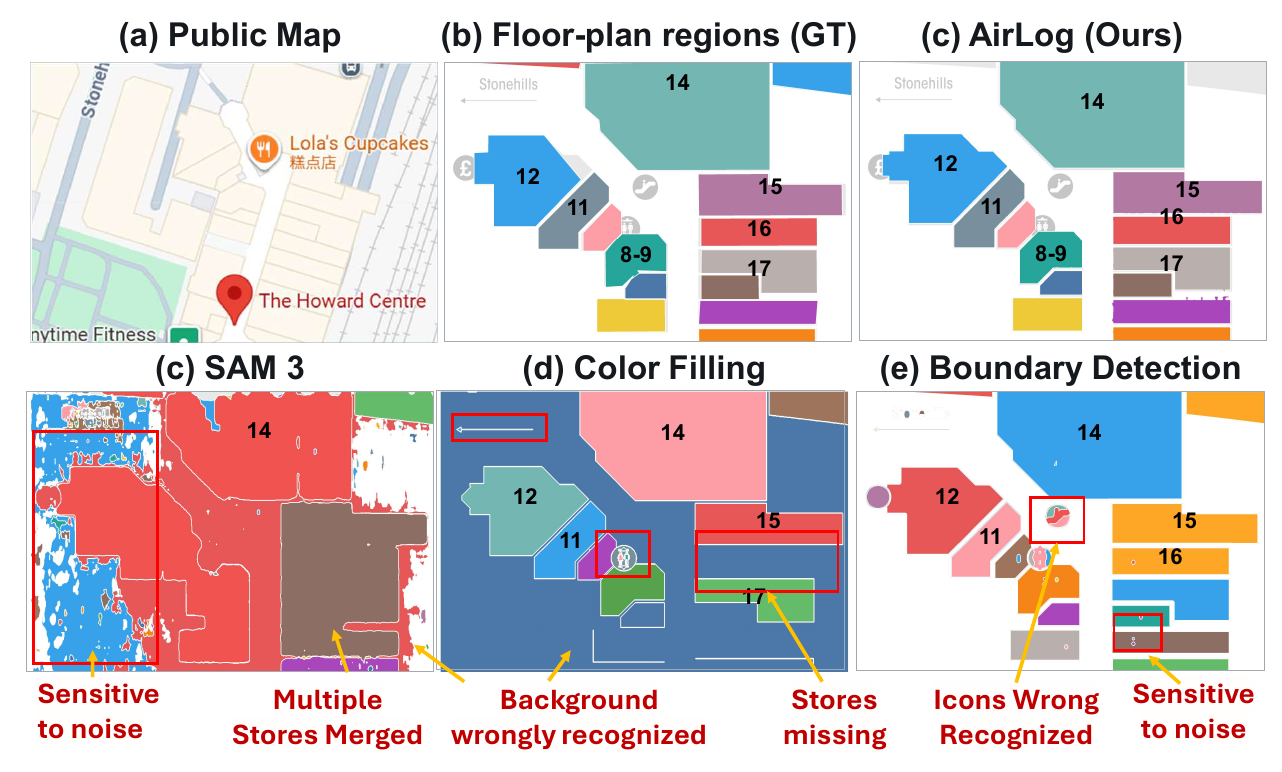}
    \vspace{-7mm}
    \caption{\textbf{From public maps to labeled region maps.}
    \textnormal{(a) The public map provides venue-level context but lacks
    floor-level region geometry. (b) The floor-plan image contains the
    annotated regions and readable labels. (c) SAM 3 absorbs background
    structures and merges adjacent regions. (d) Color Filling is distracted by
    icons and other visual structures. (e) Boundary Detection extracts text and
    noise as candidate boundaries. (f) \sysname more closely recovers the
    annotated region geometry and labels. This example is qualitative; aggregate
    results appear in Table~\ref{tab:map_parsing_results}.}}
    \label{fig:map_parsing_baseline_comparison}
    \vspace{-5mm}
\end{figure}

We begin with the map-representation bottleneck in Bob's example. His floor-plan image contains the store-level information needed to reconstruct his visit---where Uniqlo, KFC, Starbucks, and other POIs are located---but exposes this information only as pixels. As shown in Figure~\ref{fig:map_parsing_baseline_comparison}(a)--(b), a public map may identify the mall while providing no queryable floor-level POI geometry, whereas the floor-plan image visually encodes both store regions and their names. For the trajectory reconstruction in the next section, however, \sysname needs more than an image: it needs an explicit mapping from each POI label to its polygon so that downstream modules can query where stores are and reason about their spatial relationships. 

\textbf{Challenge: recovering both region geometry and readable labels.} Turning a human-facing floor-plan image into such a representation requires solving two coupled problems. First, \emph{region geometry must be recovered from a visually cluttered image}. Store boundaries are mixed with text, logos, icons, legends, decorative backgrounds, and weak separators. Second, \emph{each recovered region must be associated with the correct readable label set}. Recognizing a store name alone is insufficient: the system must know exactly which region that name describes. A mistake in either geometry or semantic association produces an incorrect map and can subsequently mislead trajectory reconstruction. 

\textbf{Why conventional and direct approaches fall short.}
The closest prior work on this problem is the shopping-mall-plan parser of
Su et al.~\cite{su2022shoppingmallplans}. Their pipeline first recognizes a
mall directory, applies threshold/edge preprocessing and two-stage region
growing to segment individual rooms, and then uses OCR-recognized room
identifiers to retrieve the corresponding room names. This work establishes
that explicitly coupling segmentation and recognition can recover both
geometry and semantics from structured shopping-mall plans.

\sysname advances this line of work to more heterogeneous human-facing
floor-plan images, where text, logos, icons, legends, reflections, weak
separators, and perspective distortion make classical region growing and
OCR-based matching brittle. Rather than requiring an OCR-readable room
identifier that can be matched back to a directory, \sysname canonicalizes the
visual layout into explicit region geometry and then assigns semantics directly
to each recovered region with a per-region semantic subagent.

Other direct alternatives expose complementary limitations. Classical
color-filling methods can merge neighboring stores or retain non-POI
structures when colors and backgrounds vary, while edge detection can mistake
text strokes and decorations for region boundaries. Generic segmentation
models such as SAM 3~\cite{carion2025sam3} face a different mismatch:
they are designed primarily for object-like regions in natural images rather
than tenant-scale graphic regions, and can therefore absorb background
structures or merge adjacent POIs. Semantic extraction alone does not solve
the problem either. Standalone OCR can miss small, stylized, rotated, or
low-resolution labels, and even a correctly recognized string does not specify
which region polygon it belongs to. A full-map VLM faces a related coupling problem:
it must recognize every POI, predict each label's position, and bind that
position to a region. A correctly recognized name can therefore still be
omitted, misplaced, or attached to the wrong polygon.
Figure~\ref{fig:map_parsing_baseline_comparison} illustrates representative
geometry failures. Appendix~\ref{sec:appendix_map_failures} pairs two
exact-region OCR failures in Figure~\ref{fig:appendix_map_ocr_examples} and
places the coordinate-binding and geometry-to-semantics mechanisms together in
Figure~\ref{fig:appendix_map_mechanisms}, beside the claims that they support.

\begin{figure}[t!]
    \centering
    \includegraphics[width=\linewidth]{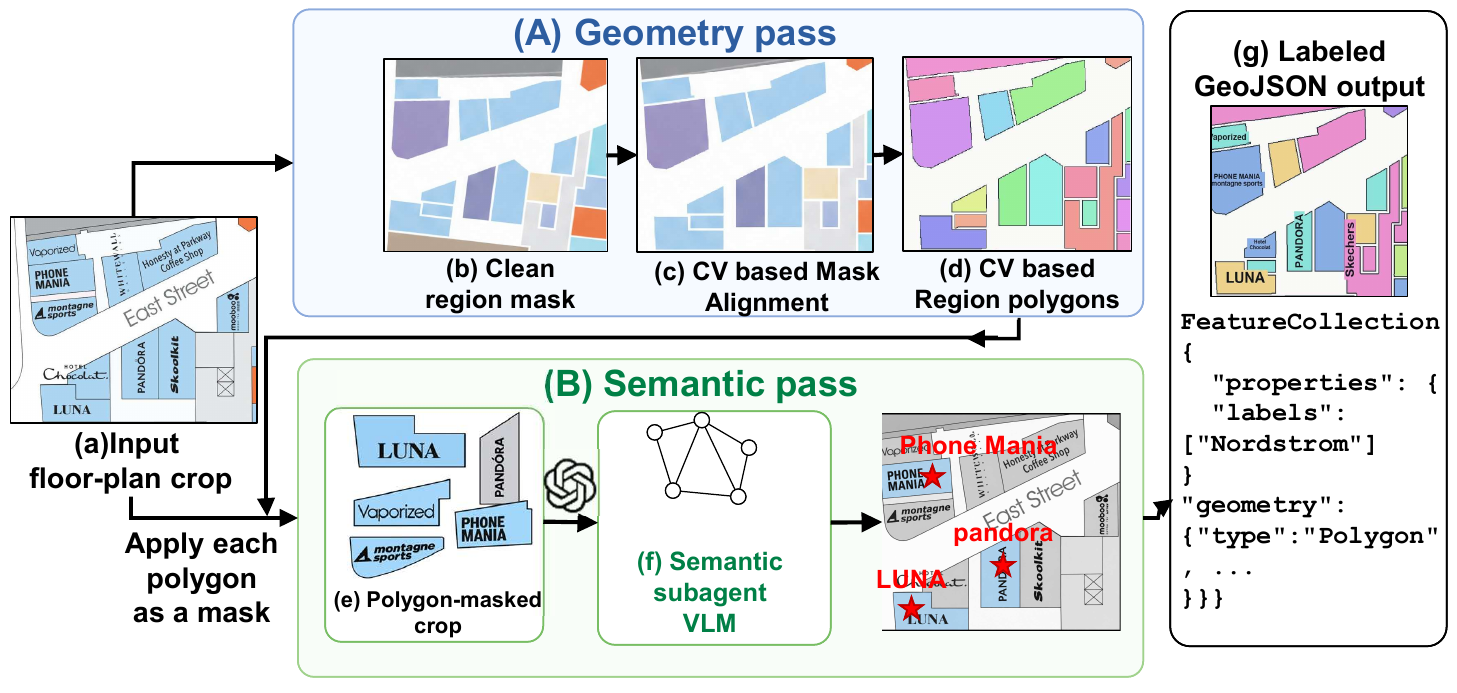}
    \vspace{-7mm}
    \caption{\textbf{Geometry-first map-parsing pipeline.}
    \textnormal{From (a) an input floor-plan crop, \sysname generates
    (b) a clean region mask, performs (c) mask alignment, and extracts
    (d) region polygons. Each polygon produces (e) a polygon-masked crop for
    (f) the per-region semantic subagent. The returned label set is written directly
    to its source feature, producing (g) a labeled region map in GeoJSON.}}
    \label{fig:map_agent_architecture}
    \Description{The pipeline first recovers region polygons from a generated
    and aligned clean mask, then invokes the per-region semantic subagent on one
    polygon-masked crop at a time and writes each label set directly to its source
    GeoJSON feature.}
    \vspace{-5mm}
\end{figure}

These observations motivate a geometry-first, sequential factorization.
Instead of asking one model to jointly recover region boundaries, recognize
readable labels, predict label coordinates, and bind labels to regions, \sysname first
recovers explicit region polygons and then uses each polygon to define one
constrained semantic task. As shown in
Figure~\ref{fig:map_agent_architecture}, the geometry stages in (a)--(d)
produce a geometry-only region map serialized as GeoJSON. Each recovered polygon is then projected
onto the original floor-plan image to produce the polygon-masked crop in (e). The
per-region semantic subagent processes one crop at a time in (f), and its
returned label set is written directly to the same GeoJSON feature in (g). The
current pipeline therefore predicts no label coordinate and requires no
separate polygon--label binding stage.

\subsection{Geometry Pass: Recovering Regions}
The geometry pass addresses the first question: \emph{where are the regions?} Directly extracting boundaries from
Figure~\ref{fig:map_agent_architecture}(a) is unreliable because region
geometry is entangled with text, logos, icons, and background variation.
\sysname therefore first canonicalizes the map's visual appearance before
extracting precise geometry. Starting from the input crop in (a), \sysname uses
\textit{GPT Image 2}~\cite{openai2026chatgptimages} to generate the clean region
mask in (b). The model redraws floor-plan regions as solid regions over a
uniform background while suppressing text, icons, legends, and other visual
clutter. The generated colors carry no POI semantics; they only make individual
spatial regions explicit. This use of image generation is intentional: the
generative model handles heterogeneous appearance, while deterministic
processing remains responsible for spatial precision.

The generated mask cannot be used directly because image generation may
introduce a small global translation or scale change relative to the source floor-plan image. \sysname therefore aligns structural gradients between the generated
mask and source image under a restricted transformation consisting of uniform
scaling and horizontal and vertical translation, producing (c). Rotation and
non-rigid deformation are excluded, and the system falls back to the identity
transformation when reliable alignment cannot be established. Deterministic CV
then separates the uniform background from foreground blocks, extracts
4-connected components, traces exterior contours and holes, and converts them
into valid \texttt{Polygon} or \texttt{MultiPolygon} features. Candidates with
insufficient support in the original image are removed. The output in (d) is a
geometry-only map that explicitly represents the floor-plan regions but does not yet
label them.

\vspace{-3mm}

\subsection{Per-Region Semantic Subagent}
Once the region polygons are fixed, \sysname uses them to constrain semantic
labeling. Here, \emph{per-region} describes the invocation unit: the subagent
processes one recovered region at a time, while the region polygon is used
only to construct its input crop. For each polygon, the system rasterizes its
exact geometry over the
original floor-plan image, retains source pixels inside the polygon, suppresses
pixels outside the polygon and inside any holes, and crops the result to the
polygon's bounding box. This produces one polygon-masked crop per GeoJSON
feature while preserving the printed POI name and excluding neighboring labels,
legends, and decorations.

The current system invokes this GPT-4o semantic subagent once for each
polygon-masked crop. Under the shared visible-label policy, the subagent returns
\texttt{{"labels":[...]}} containing all readable strings in the target
region. Because each crop is generated from a known polygon, the returned
labels are written directly to that same feature. In contrast to the earlier
full-map baseline, the current method predicts no label coordinates and
requires no point-in-polygon, nearest-neighbor, or learned cross-region binding
stage.

\vspace{-1mm}

\subsection{Generate GeoJSON}
After all recovered regions are labeled, \sysname exports a GeoJSON
\texttt{FeatureCollection}. Each feature contains \texttt{Polygon} or \texttt{MultiPolygon} geometry, along with a canonical label field, \texttt{properties.labels}, containing the recognized strings. The parser
does not emit \texttt{name}, \texttt{category}, \texttt{confidence}, corridor
polygons, walkable areas, traversability, or connectivity fields.
The result is a machine-queryable labeled region map. For Bob's example,
Uniqlo, KFC, and Starbucks become labeled region polygons that later modules
can query directly. A separate deterministic map-preparation stage derives the
corridor mask and navigability graph used in the next section; these downstream
artifacts are not part of the parser's GeoJSON output.

\section{Semantic Trajectory Reconstruction}
\label{sec:spatial_fusion}

Given the labeled region map obtained from \S\ref{sec:floorplan}, \sysname next
reconstructs the user's store-level trajectory from Wi-Fi and inertial sensing.
Rather than asking an agent to directly predict locations, in \sysname we propose a multi-agent pipeline (\S\ref{ss:multi-agent_pipeline}) that coordinates semantic Wi-Fi anchoring (\S\ref{ss:wifi_agent}), PDR-based motion
estimation (\S\ref{ss:motion_agent}), and map-constrained inference (\S\ref{sec:fusion_review}) to construct a trajectory that is both
semantically plausible and physically feasible.
\begin{figure}[t!]
    \centering
    \includegraphics[width=\linewidth]{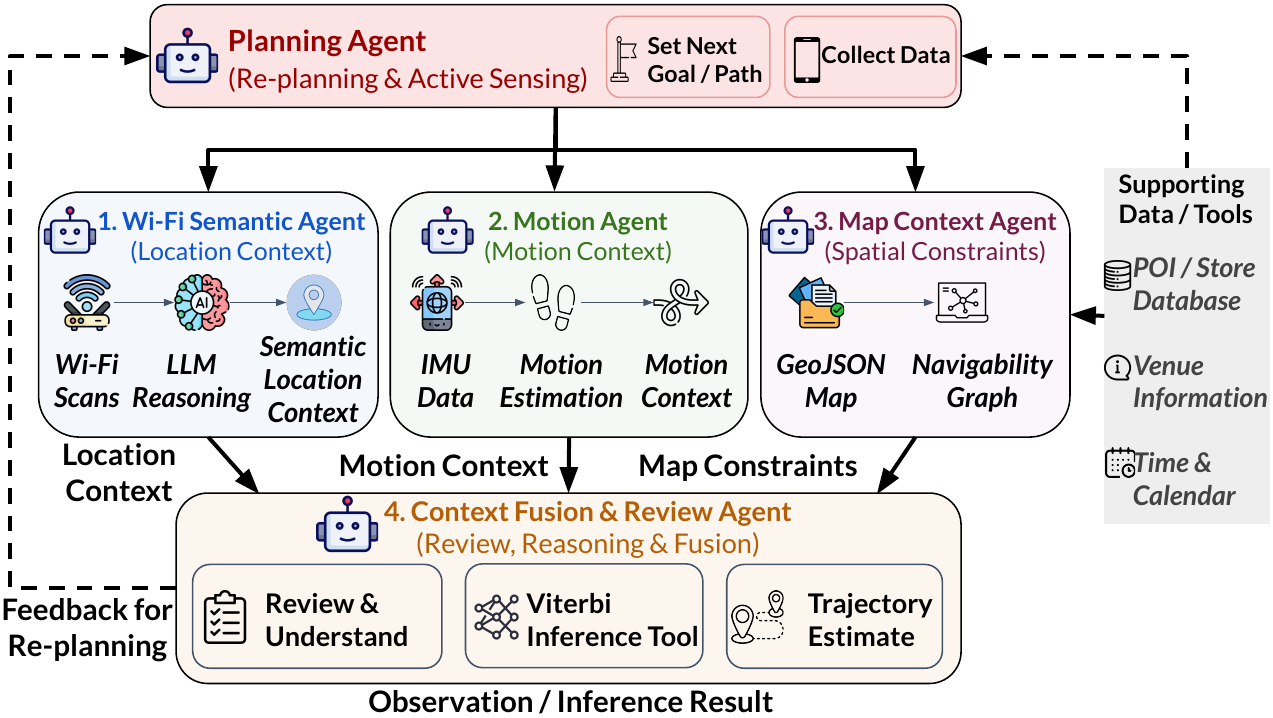}
    \vspace{-8mm}
    \caption{\textbf{Trajectory Reconstruction in \sysname.} \textnormal{The framework organizes three complementary contexts for indoor 
trajectory recovery: Wi-Fi Semantic Agent derives semantic location context, 
Motion Agent derives motion context, and Map Context Agent derives 
GeoJSON map constraints. The Context Fusion \& Review Agent integrates 
these inputs and invokes structured inference to recover a geometrically 
consistent, store-level trajectory.}}
    \label{fig:architecture}
    \vspace{-5mm}
\end{figure}

\subsection{Agent-Orchestrated Trajectory Reconstruction}
\label{ss:multi-agent_pipeline}

Figure~\ref{fig:architecture} illustrates the agent-orchestrated
context-fusion pipeline. Before processing each episode, a
\emph{Planning Agent} selects and schedules the required modules in a
ReAct-style loop~\cite{yao2023react}, revising the plan when evidence is
missing or conflicting. \sysname organizes the resulting evidence into
three complementary context streams: \emph{Wi-Fi semantic context} for
location candidates, \emph{motion context} for relative displacement and
heading changes, and \emph{map context} for physically admissible
transitions.
A structured fusion decoder combines these contexts to recover a globally
consistent trajectory, while a review stage detects failures and triggers
targeted replanning when necessary.

\subsection{Wi-Fi Semantic Agent}
\label{ss:wifi_agent}

\textbf{Validating SSID semantics in the wild.}
Prior work has shown that Wi-Fi SSIDs can encode location and service
information~\cite{seneviratne2015ssids}. We re-evaluate their usefulness in current commercial environments using Wi-Fi observations collected across six major cities (Table~\ref{tab:dataset_stats}).
Associating business-class SSIDs with nearby OpenStreetMap POIs yields 185,140 candidate SSID--POI pairs. As summarized in Table~\ref{tab:ssid_similarity_distribution}, SSID semantics remain observable at scale, but lexical matching alone is too sparse and noisy to serve as a direct signal (representative failure cases in Table~\ref{tab:bad_cases}).
We therefore treat SSIDs as \emph{weak semantic evidence} and jointly reason over SSID strings, RSSI rank, POI names, and map context. We further verify that SSID distributions are venue- and floor-specific using cross-session observations from a public benchmark~\cite{locationcompetition2020indoor}, spanning over 5.5 hours of data across 2 venues and 14 floors. Simple SSID-set Jaccard matching achieves 87.88\% accuracy in identifying the correct site--floor pair, suggesting that SSIDs provide a useful cue for automatic floor identification without site-specific surveys.

\subsubsection{SSID Semantic-noise filtering.}
Although SSID strings provide useful place-level cues, raw Wi-Fi scans contain
substantial semantic noise that can mislead the anchoring agent and inflate the
LLM context. We make
two observations from our traces. 
\begin{itemize}[leftmargin=*]
    \item \textit{Generic infrastructure SSIDs (e.g., T-mobile, China-mobile) are
prevalent but they provide less location information.} Many visible SSIDs correspond to carrier
hotspots, venue-wide guest networks, or router-default names. These networks
provide little store-specific evidence and would increase LLM input length, so
we remove them with a keyword blocklist. 
\item \textit{The strongest RSSI is not
necessarily the nearest or most useful POI cue.} Indoor RSSI is unstable under
multipath propagation, wall attenuation, crowd blockage, AP placement, and
transmit-power variation~\cite{harle2013survey,zafari2019survey}.
\end{itemize}
Therefore,
\sysname does not assign location using the strongest SSID alone. After filtering,
we aggregate scans into sliding windows, represent each SSID by its maximum RSSI
within the window, and pass the top-ranked SSID-RSSI pairs to the LLM as soft
semantic evidence. This allows the model to reason over multiple possible store
cues instead of relying on a single nearest-signal rule. Empty windows are
dropped, and gaps longer than 600\,s are flagged to avoid incorrect PDR
integration across disconnected sessions.

\subsubsection{LLM-based semantic anchoring.} 
After semantic-noise filtering, each Wi-Fi window is represented by a compact
set of top-ranked SSID--RSSI pairs. \sysname then converts this evidence into
store-level semantic anchors. We formulate SSID-to-POI grounding as a language
understanding task~\cite{korelivc2025sellma}: the model must interpret noisy
SSID strings, abbreviations, guest-network names, and RSSI ranks in the context
of the venue's POI list and floor-plan topology. For each window, \sysname
constructs a structured prompt containing three inputs: the venue's POI names,
map-derived spatial descriptions, and the filtered top-$N$ SSID--RSSI pairs.
The prompt instructs the model to treat RSSI as relative proximity evidence
rather than a deterministic distance measurement; the full prompt is shown in
Appendix~\ref{app:wifi_semantic_prompt}.

\begin{figure}[t!]
    \centering
    \includegraphics[width=\linewidth]{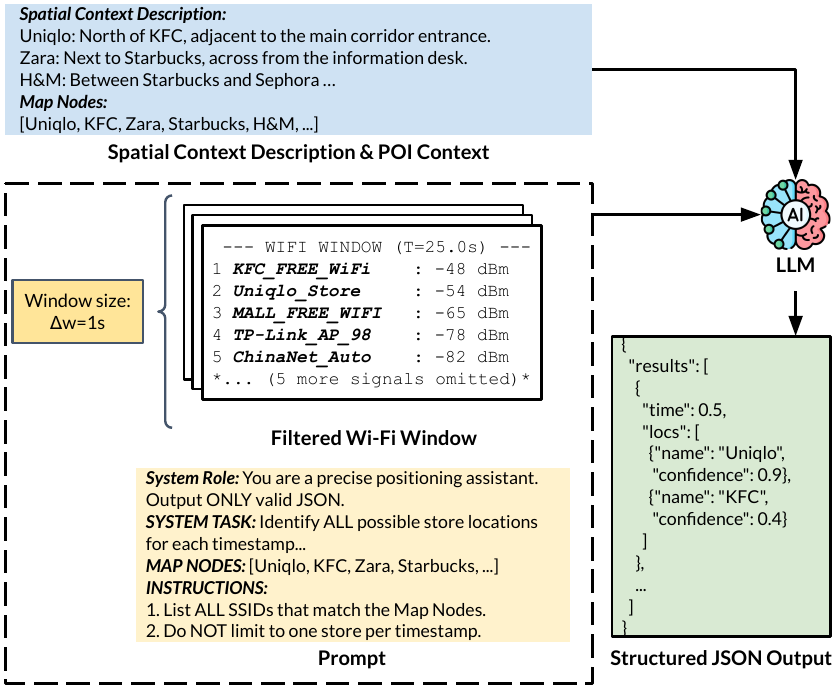}
    \vspace{-8mm}
    \caption{\textbf{LLM-based semantic anchoring pipeline.} \textnormal{Wi-Fi RSSI observations, spatial context, and POI context are combined into a structured prompt, producing multi-label POI hypotheses with confidence scores.}}
    \label{fig:semantic_anchor_pipeline}
    \vspace{-8mm}
\end{figure}

To provide spatial grounding, we augment each POI with a natural-language
descriptor derived from the navigability graph. The descriptor summarizes the
POI's relative position and nearby stores, e.g., ``Uniqlo is north of KFC and
adjacent to the main corridor entrance,'' as illustrated in
Figure~\ref{fig:semantic_anchor_pipeline}. These relational cues help the model
disambiguate SSIDs that are lexically similar or spatially overlapping, and
prevent it from treating all name matches as equally plausible. Rather than
forcing a single location decision, the model outputs all plausible nearby stores
with confidence scores between zero and one.

The resulting predictions are post-processed into semantic anchors. Each
predicted POI name is matched to the corresponding GeoJSON feature and mapped to
its polygon centroid or store region in the navigability graph. Generic outputs
such as floor identifiers, restroom tokens, and corridor labels are discarded.
Each surviving tuple $(t, n, c)$, consisting of timestamp $t$, store name $n$,
and confidence $c$, becomes a probabilistic claim that the user was near store
$n$ at time $t$. These sparse anchors provide semantic location evidence, which complements the continuous but drift-prone PDR estimate used next.

\subsection{Motion Context via PDR}
\label{ss:motion_agent}

Wi-Fi semantic anchors provide place-level cues but are sparse and
intermittent, leaving gaps between valid observations. \sysname bridges
these gaps with continuous relative motion from the phone's inertial
sensors using a standard PDR
pipeline~\cite{harle2013survey,kwapisz2011activity,weinberg2002using}: step
detections and rotation-vector-derived headings are integrated into
relative displacement and heading change between consecutive sensing
windows.

Wi-Fi anchors provide semantic positions, while PDR provides continuous motion between them. The earliest high-confidence anchor can initialize the PDR path when no surveyed starting point is available. Residual heading errors are corrected during fusion using well-separated anchors. PDR remains a motion context rather than a store-level trajectory because its errors accumulate over time~\cite{ni2022experience}.

\subsection{Context Fusion \& Review Agent}
\label{sec:fusion_review}

\begin{figure}[t!]
    \centering
    \includegraphics[width=0.9\linewidth]{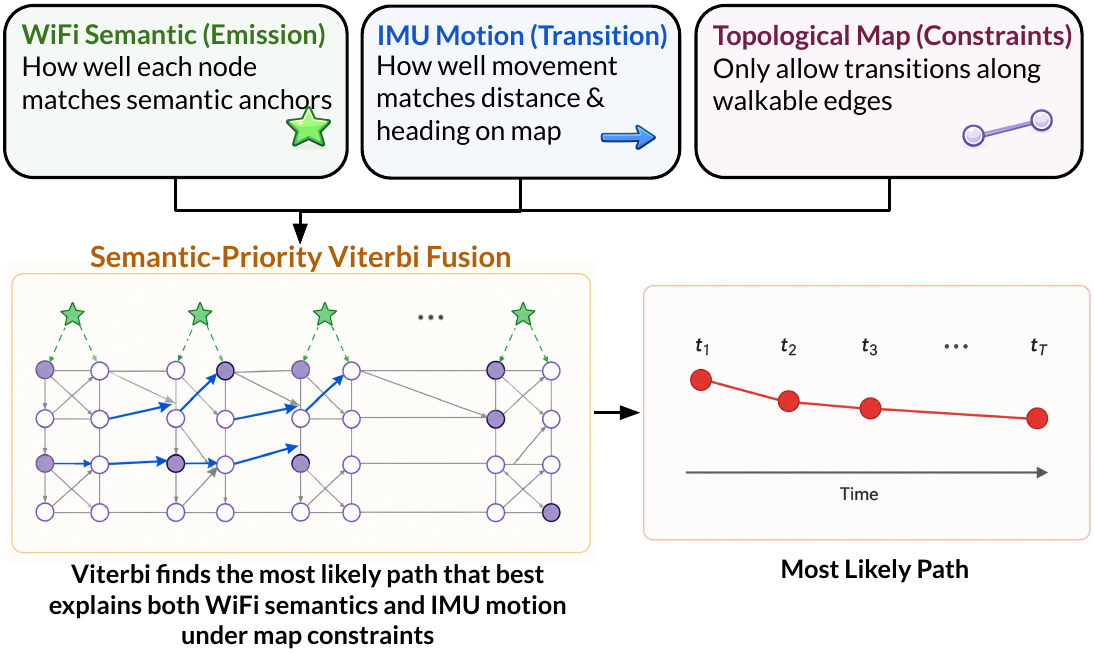}
    \vspace{-4mm}
    \caption{\textbf{Semantic-priority Viterbi fusion.} \textnormal{Wi-Fi semantic anchors score where the user is likely to be, PDR motion primitives score how the user could move, and the GeoJSON graph restricts the search to physically walkable transitions. The decoder first estimates a globally consistent trajectory, which is then reviewed by the multimodal feedback agent.}}
    \label{fig:fusion_algorithm}
    \vspace{-8mm}
\end{figure}

Prior Wi-Fi IMU floor-plan fusion systems can learn dense location histories by training fusion models over collected traces~\cite{herath2021fusion}. \sysname targets a different setting: wild indoor venues where no site-specific fingerprint survey or training trajectory is available. We therefore use a \emph{semantic-priority Viterbi} decoder that performs explicit probabilistic optimization over the GeoJSON-derived navigability graph $\mathcal{G}$~\cite{forney1973viterbi,rabiner1989tutorial}, as shown in Figure~\ref{fig:fusion_algorithm}. The decoder combines three scores. Wi-Fi semantic anchors provide the emission term: nodes near a high-confidence candidate store receive higher likelihood. PDR provides the transition term: candidate graph edges are favored when their distance and heading agree with the observed displacement and turn angle. The map provides the hard topological constraint: Viterbi only considers adjacent nodes in $\mathcal{G}$, which restricts the search to physically walkable routes. We summarize the decoded trajectory as
\begin{align}\footnotesize
    \hat{S}_{1:T} =
    \arg\max_{S_{1:T}\in\mathcal{G}}
    \sum_t
    \big[
        \beta E_t(S_t) +
        \alpha A_t(S_{t-1},S_t) \nonumber\\
        {}+
        \gamma D_t(S_{t-1},S_t) +
        C_t(S_{t-1},S_t)
    \big],
\end{align}
where $E_t$ is semantic consistency with Wi-Fi anchors, $A_t$ and $D_t$ are IMU heading and distance consistency, and $C_t$ is an anchor-based heading-calibration bonus when consecutive high-confidence semantic anchors imply a reliable direction. Concretely, for a candidate transition from node $u$ to node $v$, with edge direction $\theta(u,v)$, edge length $\ell(u,v)$, IMU motion primitive $(d_t,\Delta\phi_t)$, and active semantic anchors $\hat{L}_t=\{(n_j,c_j)\}$, these terms are
\begin{align}\footnotesize
    E_t(v) &=
    \sum_{(n_j,c_j)\in\hat{L}_t}
    c_j\left(-\frac{\operatorname{dist}(v,\mathcal{P}_{n_j})^2}{2\sigma^2}\right),\\
    A_t(u,v) &=
    \log\max\left(\cos(\theta(u,v)-\Delta\phi_t),\,\epsilon\right),\\
    D_t(u,v) &=
    -\left(d_t-\ell(u,v)\right)^2 .
\end{align}
Here $\mathcal{P}_{n_j}$ denotes the polygon or centroid of the POI named by anchor $n_j$, and $\sigma$ controls the spatial bandwidth of semantic evidence. In short, Wi-Fi estimates where the user is likely to be, IMU estimates how the user moved, and the map defines where the user can move.

\textbf{Review and Feedback Loop}:  This optimization produces an initial trajectory estimate, which \sysname then evaluates using diagnostic signals such as position jumps, stuck segments, and PDR-to-map scale drift. If these indicate a low-quality decode, a deterministic grid search re-runs Viterbi over preset $(\alpha,\beta,\gamma,\text{calib\_weight})$ configurations and keeps the lowest-variance result; if issues persist, the Planning Agent inspects the flagged windows using read-only Wi-Fi, anchor, spatial, and PDR context and selects one bounded corrective action, such as re-running semantic anchoring with adjusted parameters or known SSID mappings, reweighting the fusion decoder within a fixed valid range, falling back to substring-based Wi-Fi matching, patching a single behavior label, or accepting the result. Only Wi-Fi anchoring and fusion may be revisited, each within a small retry budget, while map parsing and PDR are fixed; the investigation itself is capped by a round limit and a wall-clock budget, after which the best result found is used. The Planning Agent's decision calls use a low sampling temperature, so repeated runs can select different parameter values within these bounded ranges rather than always reaching an identical trajectory.

\section{Journal Generation}
\label{sec:diary_generation}

\begin{figure}[t!]
    \includegraphics[width=\linewidth]{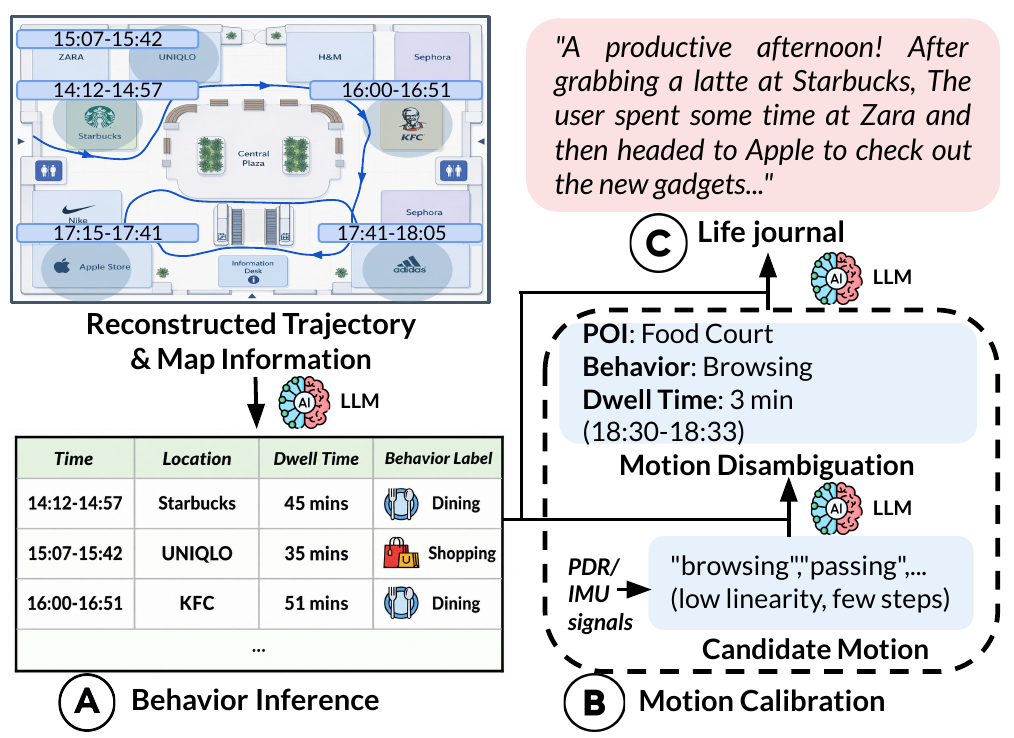}
    \vspace{-8mm}
    \caption{\textbf{Workflow of Diary Generation.} \textnormal{The reconstructed
trajectory is aggregated into store-level visits by a behavior inference module
(A). A motion calibration module (B) then refines each activity label using PDR
geometry and location context. Finally, a diary agent (C) synthesizes the
calibrated behavior sequence into a third-person Life Journal.}}
    \label{fig:journal}
    \vspace{-5mm}
\end{figure}

Spatial fusion produces a corrected trajectory, but the result remains a
localization sequence rather than a human-readable account. As shown in
Fig.~\ref{fig:journal}, \sysname transforms this sequence into a diary through
three stages: behavior inference (A), motion calibration (B), and journal
generation (C).

\subsection{Behavior Inference}

As shown in Fig.~\ref{fig:journal}(A), \textbf{behavior inference} converts the
fused trajectory into store-level visits. For each POI, \sysname merges
consecutive contacts inside or near the store region, producing an enter time,
exit time, dwell duration, semantic confidence, and local motion summary.
Consecutive contacts are merged using a trajectory-adaptive temporal tolerance, allowing segmentation to adapt to different sampling rates. Short contacts are treated as passing behavior, whereas longer
dwell periods form candidate visits. Structural POIs such as corridors,
elevators, and restrooms are excluded regardless of dwell time.

For each candidate visit, the inference agent uses the store name and category,
dwell duration, neighboring POIs, and an IMU-derived activity cue to infer the
most likely behavior. The motion cue can come from either a conventional
smartphone activity recognizer or a zero-shot activity model
~\cite{kwapisz2011activity,shoaib2014fusion,10590466}. The output is a compact
behavior sequence containing the store, time interval, dwell duration, inferred
action, and confidence.

\subsection{Journal Generation}

\noindent\textbf{Motion Calibration.}
As illustrated in Fig.~\ref{fig:journal}(B), motion calibration refines
candidate activity labels using PDR-derived motion features and location
context. Step count, displacement, path linearity, and mean speed first
produce a candidate label from
${\text{stationary, walking, browsing, transit}}$, which an LLM then
disambiguates using the store category, dwell time, and neighboring POIs.
 Before narration, an
evidence-filtering step removes unreliable WiFi anchors. Specifically, a
POI's WiFi evidence is discarded when its active window exceeds a predefined
fraction of the visit while scan density remains low, indicating repeated
background detections rather than a concentrated visit. This rule overrides
even high-confidence WiFi anchors.

\noindent\textbf{Journal Generation.}
As shown in Fig.~\ref{fig:journal}(C), journal generation transforms the
validated visits and calibrated actions into a third-person Life Journal.
The diary agent first constructs a structured outline containing the
chronological flow, dominant activity, inferred purpose, and salient visit
characteristics, and then rewrites the outline as a fluent narrative. This
two-stage process keeps the journal grounded in validated behaviors while
allowing natural transitions and varied phrasing.
\begin{table}[t!]
\centering
\caption{\textbf{Map-parsing results on the complete benchmark.}
\textnormal{Higher is better. The two \sysname rows share identical region
geometry and differ only in semantic-labeling configuration.}}
\label{tab:map_parsing_results}
\vspace{-3mm}
\footnotesize
\setlength{\tabcolsep}{3.0pt}
\rowcolors{2}{gray!10}{white}
\begin{tabular}{@{}lccc@{}}
\toprule
\textbf{Method} & \textbf{Area F1} $\uparrow$ &
\shortstack{\textbf{Region}\\\textbf{Purity}} $\uparrow$ &
\shortstack{\textbf{Matched}\\\textbf{Label F1}} $\uparrow$ \\
\midrule
Color Filling                         & 0.277 & 0.510 & 0.523 \\
Su et al.-style Boundary Detection   & 0.144 & 0.468 & 0.506 \\
SAM 3                                 & 0.037 & 0.306 & 0.155 \\
\midrule
\sysname w/o semantic subagent       & \textbf{0.705} & \textbf{0.767} & 0.734 \\
\textbf{\sysname w/ semantic subagent} & \textbf{0.705} & \textbf{0.767} & \textbf{0.814} \\
\bottomrule
\end{tabular}
\vspace{-4mm}
\end{table}

\vspace{-1mm}
\section{Implementation}
\label{sec:implementation}

\textbf{Prototype.} \sysname is implemented as a Python prototype. A preprocessing
module normalizes timestamps, converts raw Wi-Fi records into SSID--RSSI
snapshots, resamples IMU streams for PDR, and serializes each session into a
structured JSON file. The JSON representation contains Wi-Fi windows, PDR motion
primitives, and map context including POI names, polygon centroids, and
graph-based spatial descriptors. To facilitate reproducibility, we will publicly release the implementation and
experimental configurations upon publication.

\noindent \textbf{Model backends.} We evaluate \sysname with both closed-source and open-weight LLMs. GPT-4o is
the default backbone, while GPT-5.4, Claude Sonnet 4.6, gpt-oss-120b, Gemma 4, Qwen 3, and DeepSeek-R1 are used for
cross-model evaluation~\cite{openai2026gpt4o,openai2026gpt54,anthropic2026sonnet46,
openai2025gptoss}. The map-semantic comparison additionally uses GPT-5.5
~\cite{openai2026gpt55}. Across backbones, preprocessing, graph construction,
Viterbi parameters, prompts, and post-processing remain fixed.

\noindent \textbf{Prompting and tools.} \sysname uses a lightweight ReAct-style agent loop~\cite{yao2023react},
with LLMs handling semantic reasoning and deterministic Python modules
performing validation and spatially constrained trajectory fusion. Agent
outputs follow structured JSON schemas for semantic anchoring, behavior
inference, and diary generation.

\noindent \textbf{Public dataset.} We use the sample data from Indoor Location Competition 2.0
~\cite{locationcompetition2020indoor}, which provides smartphone sensor
traces, Wi-Fi scans, floor-plan images, and GeoJSON maps from two large
shopping malls in Hangzhou. Site~1 covers five floors with 642 traces
totaling 563.8 minutes, while Site~2 covers nine floors with 429 traces
totaling 290.8 minutes. Overall, the dataset contains 1,071 traces,
19,712 Wi-Fi scans, and 8,704 surveyor-labeled waypoints. Each trace
provides time-synchronized sensor streams and surveyed waypoints with
absolute coordinates, enabling quantitative evaluation of zero-shot
trajectory reconstruction in multi-floor indoor environments.

\noindent \textbf{Self-collected dataset.} We collect smartphone traces using Sensor Logger and
export the recorded data for offline processing~\cite{choi2026sensorlogger}. A total of 603.0 hour data was collected (Table~\ref{tab:dataset_stats}). For
each experiment, a volunteer starts recording before entering the venue and
carries the phone naturally while performing ordinary activities. Wi-Fi scans are
recorded at the application default interval 10~s,
and IMU streams are resampled to 50\,Hz before PDR processing. Each session is
paired with a venue floor plan and a manual reference record of visited stores
and dwell intervals. The floor plan is captured by the
volunteer as a photo of the mall directory before the session.
\sysname uses readable floor-plan images obtained from public venue websites, online map directories, or crowdsourced photos when available. We use the self-collected dataset primarily for Wi-Fi semantics analysis 
and journaling validation. 
For journaling evaluation, we recruited 11 volunteers to annotate a 
subset of the traces in Table~\ref{tab:dataset_stats} with reference 
journals, yielding a total of 33\,h\,37\,min of recorded time, 
of which 20\,h\,55\,min correspond to indoor activity.
\vspace{-5mm}
\section{Evaluation}
\label{sec:evaluation}

\subsection{Map Parsing Evaluation}
\label{subsec:map_parsing_evaluation}
\vspace{-1mm}

\begin{figure*}[t!]
\centering
\begin{minipage}[t]{0.23\linewidth}
    \centering
    \includegraphics[width=\linewidth]{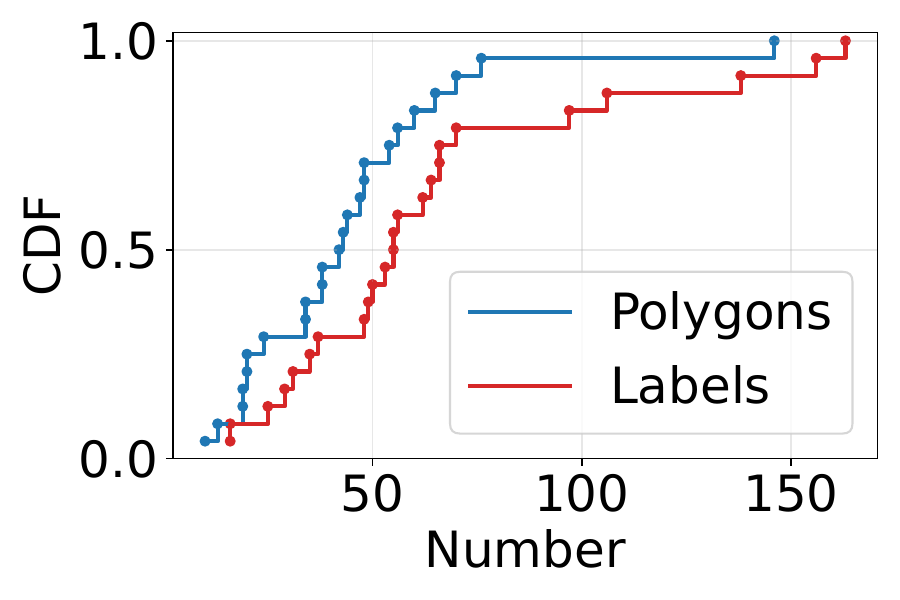}
    \vspace{-7mm}
    \caption{\textbf{Floor-plan image complexity.}}
    \label{fig:map-interference}
    \vspace{-5mm}
\end{minipage}
\hfill
\begin{minipage}[t]{0.76\linewidth}
    \centering
    \begin{subfigure}[t]{0.24\linewidth}
        \includegraphics[width=\linewidth]{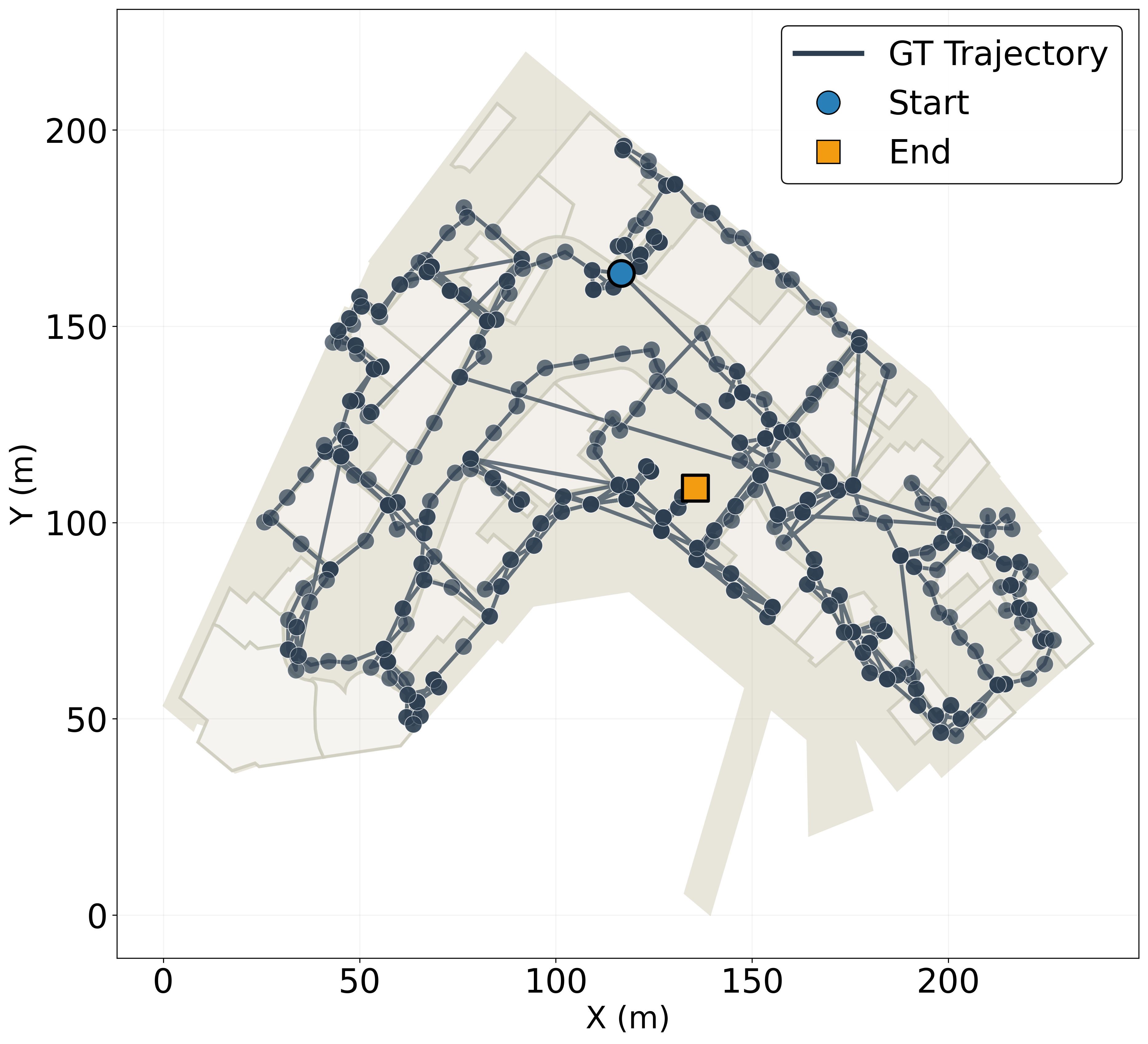}\vspace{-3mm}
        \caption{\textnormal{Ground truth}}\vspace{-1mm}
        \label{fig:waypoint_trajectory}
    \end{subfigure}
    \hfill
    \begin{subfigure}[t]{0.24\linewidth}
        \includegraphics[width=\linewidth]{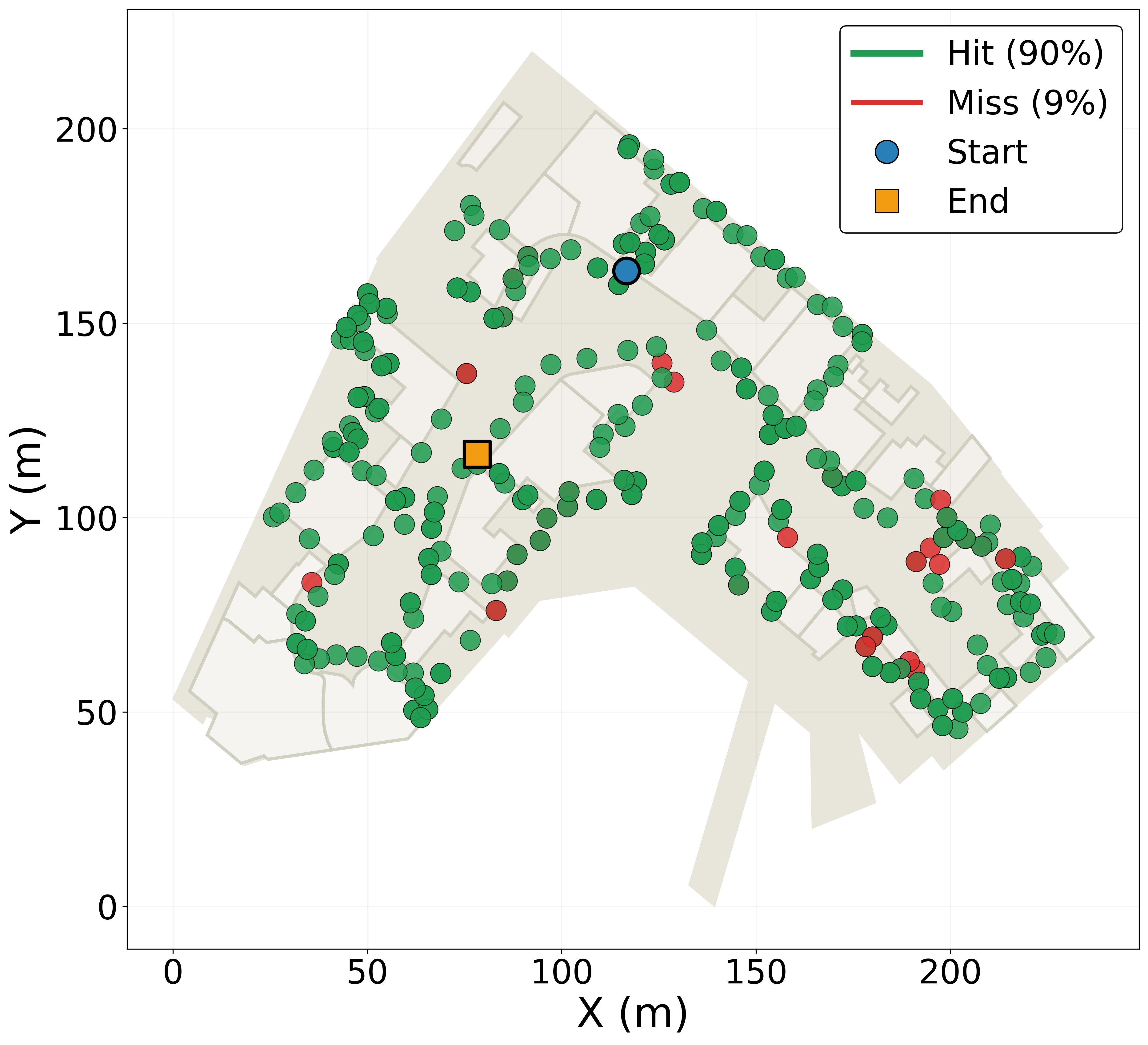}\vspace{-3mm}
        \caption{\textnormal{AirLog (Ours)}}\vspace{-1mm}
        \label{fig:airlog_subset_hit_spatial}
    \end{subfigure}
    \hfill
    \begin{subfigure}[t]{0.24\linewidth}
        \includegraphics[width=\linewidth]{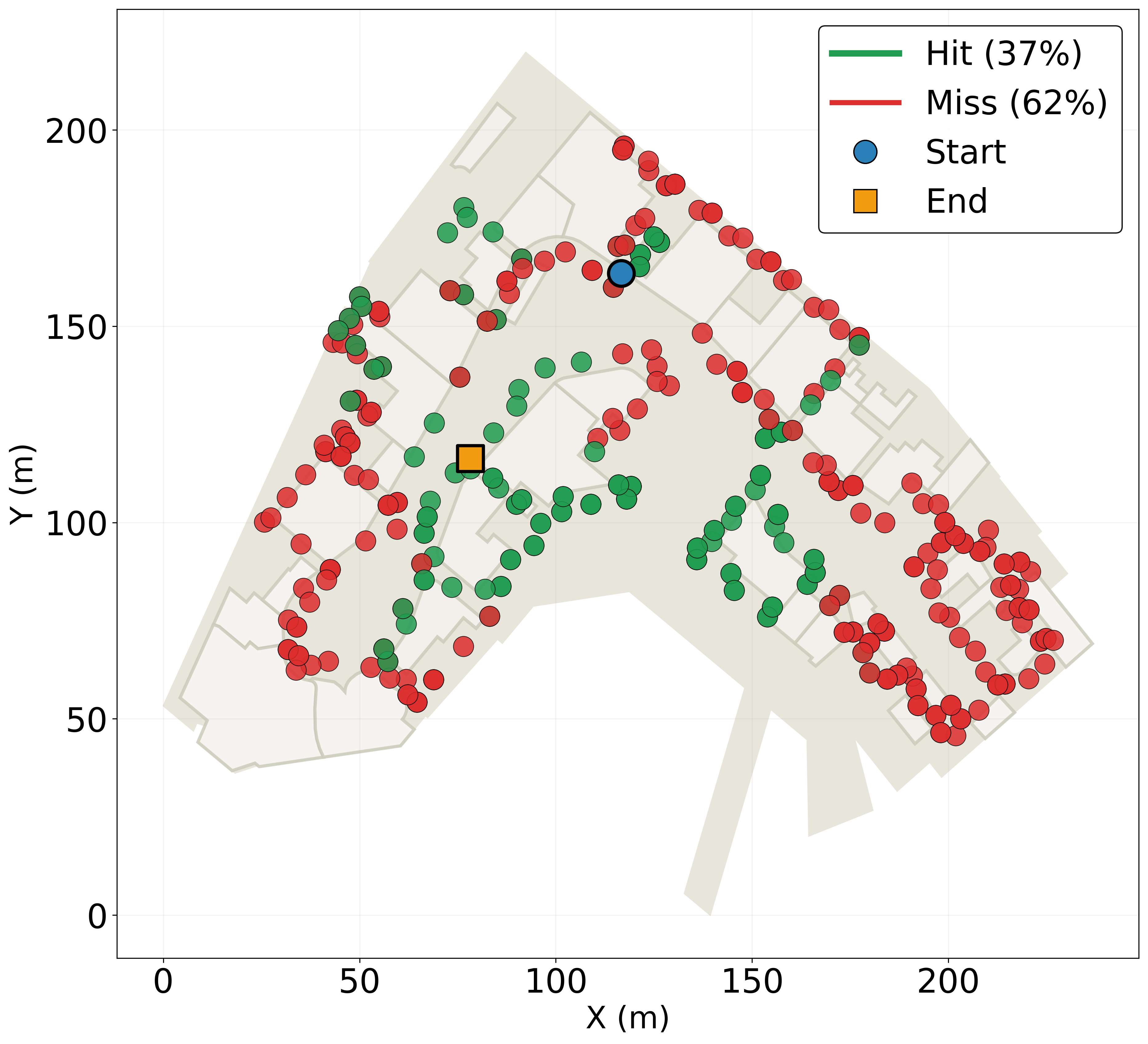}\vspace{-3mm}
        \caption{\textnormal{SenLLM}}\vspace{-1mm}
        \label{fig:senllm_subset_hit_spatial}
    \end{subfigure}
    \hfill
    \begin{subfigure}[t]{0.24\linewidth}
        \includegraphics[width=\linewidth]{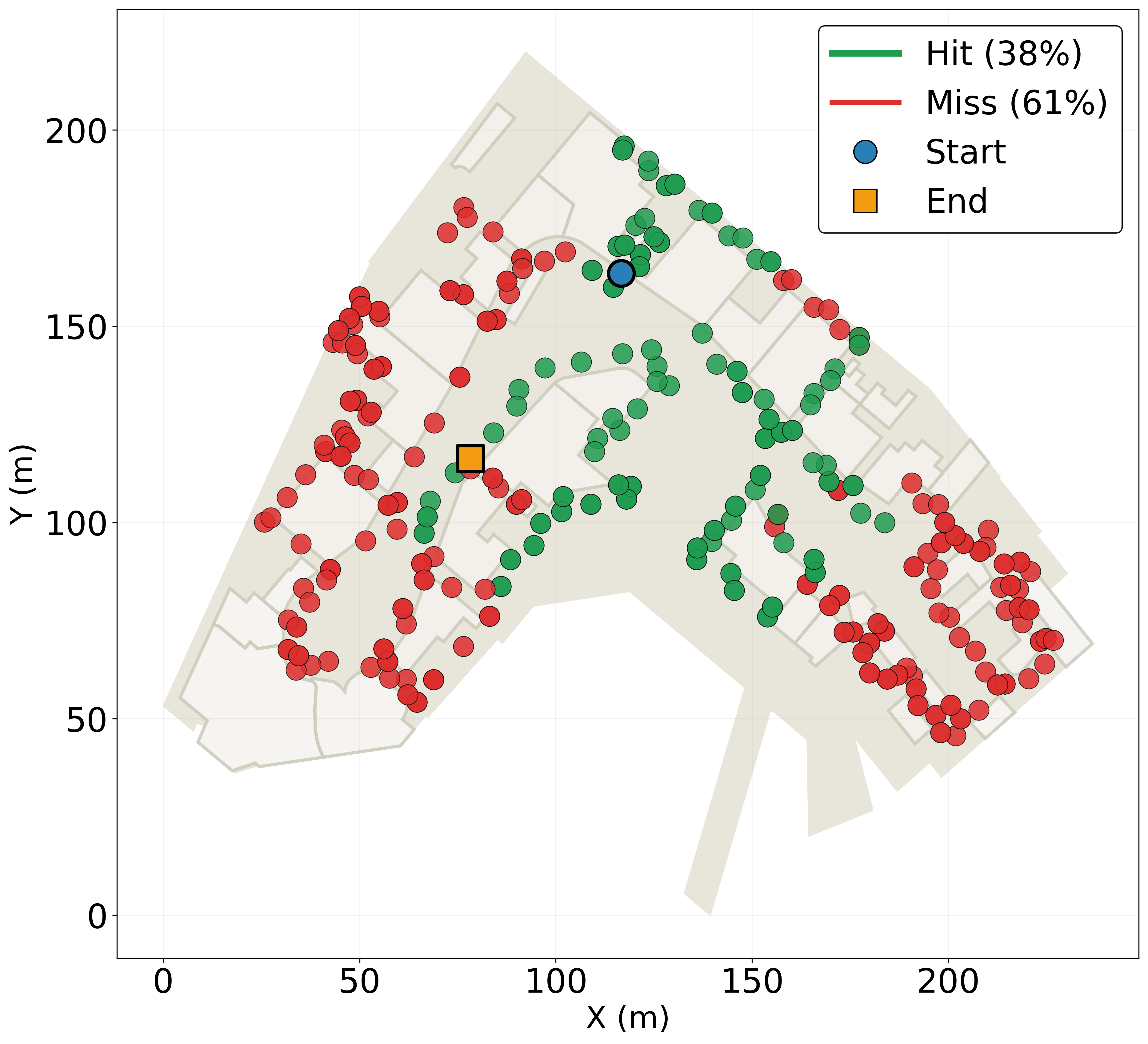}\vspace{-3mm}
        \caption{\textnormal{AutoLife}}\vspace{-1mm}
        \label{fig:autolife_subset_hit_spatial}
    \end{subfigure}
    \vspace{-3mm}
    \caption{\textbf{Spatial distribution of subset hit/miss results across localization methods.}
    \textnormal{Green dots indicate waypoints where the GT location
    falls within the predicted subset; red dots indicate misses.}}
    \label{fig:trajectory_comparison}
    \vspace{-5mm}
\end{minipage}
\end{figure*}

\noindent\textbf{Metrics.}
We report three complementary scores.
\emph{(1) Area F1} measures tolerant one-to-one region recovery, capturing whether annotated regions are successfully recovered without requiring exact boundaries.
\emph{(2) Region Purity} measures the fraction of predicted area overlapping annotated foreground, reflecting how well predictions avoid background and other clutter.
\emph{(3) Matched Label F1} evaluates semantic label accuracy after geometry-based matching between predicted and ground-truth regions.
Detailed matching rules, averaging procedures, and label normalization are provided in Appendix~\ref{sec:map-metrics}.

\noindent\textbf{Baselines.}
We compare three visual pipelines and two \sysname semantic-labeling configurations.
\emph{(1) Color Filling} groups connected regions by color and binds OCR labels by
spatial overlap. \emph{(2) Boundary Detection / Su et al.} represents the classical
edge/region-growing approach in mall-plan parsing~\cite{su2022shoppingmallplans},
with OCR applied separately to candidate regions; because no official
implementation is used, this is our implementation of the procedure described
in the paper. \emph{(3) SAM 3}~\cite{carion2025sam3} uses the fixed text-prompted
configuration documented in Appendix~\ref{sec:benchmark_snapshot}, followed by
full-crop OCR with the same spatial binding strategy as Color Filling. We also
report a uniform threshold-0.5 SAM 3 sensitivity rerun in the appendix. Since
these methods differ in both region extraction and label binding,
Table~\ref{tab:map_parsing_results} presents an end-to-end pipeline comparison
rather than a controlled comparison under a shared semantic-labeling module.

The two \sysname configurations use identical frozen predicted polygons and the
same visible-label policy. \emph{Full-map} uses GPT-5.5 to jointly predict labels
and coordinates for the entire map before binding them to polygons, whereas the
current configuration invokes GPT-4o on each polygon-masked crop and writes the
returned label list directly to the corresponding GeoJSON feature.

\noindent\textbf{Setting and map complexity.}
The complete benchmark contains 24 floor-plan images spanning both public
web-published mall directories and in-the-wild photographs. Twenty correspond
to publicly discoverable malls with official websites and public
directory/floor-plan images; the remaining four were photographed during our
in-the-wild sensing collection and may contain realistic perspective
distortion, illumination variation, and color shift. Region geometry was
manually maintained in QGIS against the exact pixel-aligned crop. GPT-5.5 was
used only as an auxiliary label-auditing tool, and all final ground-truth labels
were manually verified against the source images. All method--image evaluations
completed under evaluator v8; the archived benchmark snapshot and runtime
details are documented in Appendix~\ref{sec:benchmark_snapshot}.
Figure~\ref{fig:map-interference} summarizes the density of OCR text relative to geometry-valid annotated regions, characterizing visual clutter rather than recognition accuracy.

\noindent\textbf{Region recovery.}
\sysname substantially outperforms the visual baselines in both Area F1 and
Region Purity, indicating more complete region recovery and cleaner foreground
localization. Area F1 increases from 0.277 for Color Filling to 0.705 for
\sysname, while Region Purity increases from 0.510 to 0.767. These scores
reflect tolerant region matching and foreground coverage, respectively, rather
than exact pixel overlap. The conclusion is unchanged under standard
IoU-based metrics: \sysname obtains 0.590 Region F1@0.5 and 0.475 PQ@0.5,
compared with 0.229 and 0.199 for Color Filling; Appendix~\ref{sec:map-metrics}
reports the full sanity check.

\noindent\textbf{Semantic-labeling comparison.}
Table~\ref{tab:semantic_factorization_main} compares the two production
semantic-labeling configurations under identical frozen region geometry and the
same visible-label policy. \emph{Calls} denotes completed semantic
records/application attempts, \emph{Tokens} reports aggregate input/output
tokens, and \emph{Time} is the summed model-response latency rather than
wall-clock batch time. \emph{Label F1} is the Matched Label F1 defined above.
Because the backend model and input scope change together, this is an
operational comparison rather than a controlled ablation; we do not attribute
the difference independently to model choice or region cropping.

\begin{table}[t!]
\centering
\caption{\textbf{Semantic labeling comparison.}
\textnormal{The GPT-4o per-region run contains 1,920 completed semantic records
from 1,927 application attempts; all seven extra attempts were recovered by
retries and no record was excluded.}}
\label{tab:semantic_factorization_main}
\vspace{-3mm}
\footnotesize
\setlength{\tabcolsep}{1.7pt}
\begin{tabular}{@{}llrrrr@{}}
\toprule
\textbf{Model} & \textbf{Input} & \textbf{Calls} &
\textbf{Tokens (I/O)} & \textbf{Time (s)} &
\textbf{Label F1} \\
\midrule
GPT-5.5 & Full map
& 24/24 & 127.4k/137.0k & 1,984.0 & 0.734 \\

\textbf{GPT-4o} & \textbf{Region crop}
& 1,920/1,927 & 524.2k/24.6k & 1,815.9 & \textbf{0.814} \\
\bottomrule
\end{tabular}
\vspace{-3mm}
\end{table}

\noindent\textbf{Semantic labeling.}
With region geometry and visible-label policy fixed, the GPT-4o region-crop
configuration reaches 0.814 Matched Label F1 (precision 0.789, recall 0.842),
compared with 0.734 for full-map GPT-5.5 (precision 0.728, recall 0.740). The
0.080 difference describes the two deployed configurations only because model
and input scope change together. The GPT-4o run completed all 1,920 semantic
records; seven transient failed attempts across six polygons were recovered
within the retry budget, so no sample was excluded from evaluation.

\vspace{-1mm}
\subsection{Store-Level Trajectory \& Visit Recovery}

Next, we evaluate how accurately \sysname reconstructs a user's trajectory and recovers the stores visited along the trajectory.

\noindent \textbf{Metrics.} We evaluate recovery at two complementary levels.
For \emph{point-level trajectory recovery}, \emph{(1) Subset Hit} measures whether
the predicted set covers all ground-truth POIs, while \emph{(2) Avg Hit Ratio}
measures the fraction of ground-truth POIs covered, using the five nearest POIs
and an 8~m spatial threshold with a $\pm$5~s temporal window. These metrics are
evaluated on the public benchmark with surveyed ground truth
(\S\ref{sec:implementation}). For \emph{store-level visit recovery}, we further
evaluate self-collected diary cases (\S\ref{sec:implementation}), where
\emph{(3) Recall} measures recovered ground-truth visits and \emph{(4) SeqSim}
measures ordering consistency between predicted and ground-truth POI sequences
using normalized edit distance.

\begin{figure}[t!]
    \centering
    \includegraphics[width=\linewidth]{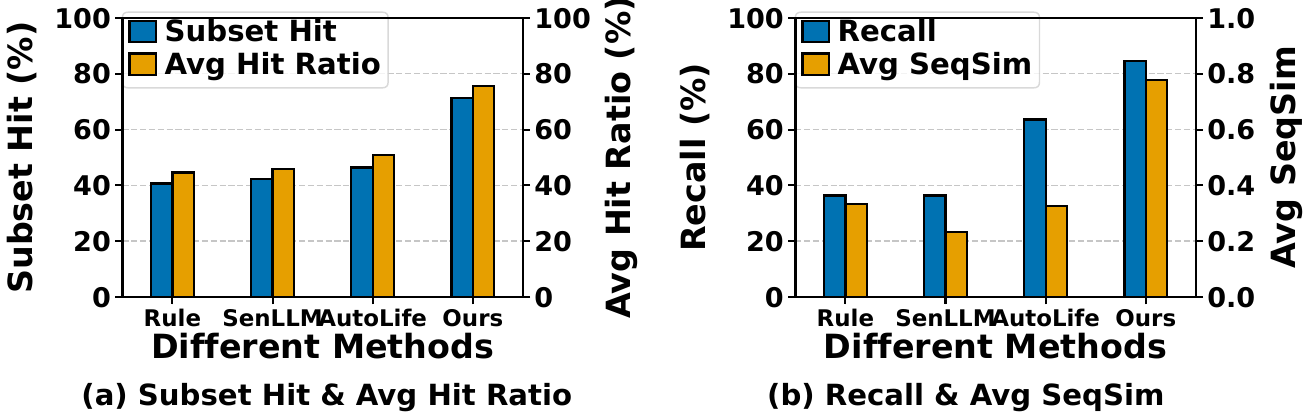}
    \vspace{-8mm}\caption{\textbf{Trajectory reconstruction evaluation.}}
    \label{fig:trajectory}
    \vspace{-6mm}
\end{figure}

\noindent \textbf{Baselines.} We compare \sysname against three baselines.
~\emph{(1) SenLLM}, inspired by Penetrative AI~\cite{xu2024penetrative}, is a
pure LLM reasoning baseline that receives the same formatted sensing context
and infers the POI sequence from motion and Wi-Fi data. ~\emph{(2) AutoLife}
\cite{xu2025autolife} is adapted to indoor settings by using the provided
floor-plan image as VLM location context instead of a GPS-queried map.
~\emph{(3) Rule-based Trajectory Reconstruction} assigns each timestamp to the
POI with the strongest nearby Wi-Fi signal.

\noindent \textbf{Results.} 
Figure~\ref{fig:trajectory} shows that \sysname outperforms all baselines, improving Subset Hit and Avg Hit Ratio over AutoLife by 24.8\% and 24.6\%, respectively. It also achieves 84.6\% Recall and 0.778 SeqSim, compared with 63.6\% and 0.327 for AutoLife. The limited gain of AutoLife over SenLLM suggests that floor-plan images alone are insufficient for reliable indoor trajectory reasoning, whereas \sysname combines map topology, Wi-Fi semantics, and motion continuity for more accurate and physically consistent reconstruction; remaining errors mainly occur in anchor-sparse corridors and dense areas with overlapping Wi-Fi semantics (Fig.~\ref{fig:trajectory_comparison}).

\begin{figure}[t!]
    \includegraphics[width=0.9\linewidth]{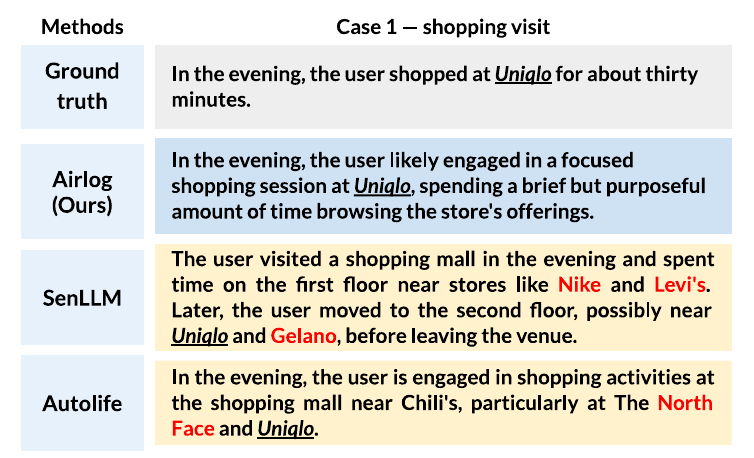}
    \vspace{-5mm}
    \caption{\textbf{Comparison of generated journals.} \textnormal{\underline{\textit{Underlined italic}} denotes ground-truth POI names. {\color{red} Red} denotes hallucinated or incorrect content. More details in the Appendix~\ref{sec:appendix_diary_examples}.}}
    \label{fig:diary_comparison}
    \vspace{-8mm}
\end{figure}

\subsection{Journalling Recovery}
\textbf{Metrics.} To evaluate the quality of generated journals, we measure their
similarity to reference journals using chrF~\cite{popovic2015chrf} and
BERTScore~\cite{zhang2020bertscore}. We also report hallucination rate, which is marked as
hallucinated if the response contains content that is not supported by
evidence and is factually inconsistent with the target context.

\noindent \textbf{Settings}: We recruit 11 volunteers to collect indoor shopping traces.
Each participant carries a smartphone during natural shopping activities while
the device records Wi-Fi scans and IMU data. After each session, the participant
writes a concise reference journal describing the stores visited, dwell periods,
and main activities.

\noindent \textbf{Results.} 
Table~\ref{tab:diary} shows that \sysname achieves the best performance
across all metrics. It reduces the hallucination rate from 0.667 for
SenLLM and 0.636 for AutoLife to 0.000, while improving chrF by 26.9\%
and 10.2\%, and BERTScore F1 by 53.3\% and 23.7\%, respectively.
Figure~\ref{fig:diary_comparison} further shows that the baselines
introduce unsupported stores, whereas \sysname generates journals
grounded in the recovered POI visit sequence.

\begin{table}[t!]
\centering
\caption{\textbf{Diary generation quality evaluation.}
\textnormal{\textit{Hall.} Rate indicates the hallucination rate,
\textit{chrF} denotes the character F-score, and \textit{P}, \textit{R},
and \textit{F1} refer to the precision, recall, and F1 score of BERTScore.}}
\label{tab:diary}
\vspace{-3mm}
\footnotesize

\rowcolors{3}{gray!8}{white}
\begin{tabular}{lccccc}
\toprule
\multirow{2}{*}{\textbf{Method}}
& \multirow{2}{*}{\textbf{Hall. Rate} $\downarrow$}
& \multirow{2}{*}{\textbf{chrF} $\uparrow$}
& \multicolumn{3}{c}{\textbf{BERTScore} $\uparrow$} \\
\cmidrule(lr){4-6}
& & & P & R & F1 \\
\midrule

SenLLM
& 0.667 & 0.331 & 0.263 & 0.403 & 0.330 \\

AutoLife
& 0.636 & 0.381 & 0.363 & 0.458 & 0.409 \\

\midrule
\textbf{\sysname}
& \textbf{0.000}
& \textbf{0.420}
& \textbf{0.485}
& \textbf{0.529}
& \textbf{0.506} \\

\bottomrule
\end{tabular}
\vspace{-5mm}
\end{table}

\begin{figure*}[t!]
\centering

\begin{minipage}[t]{0.24\linewidth}
\centering
\includegraphics[width=\linewidth]{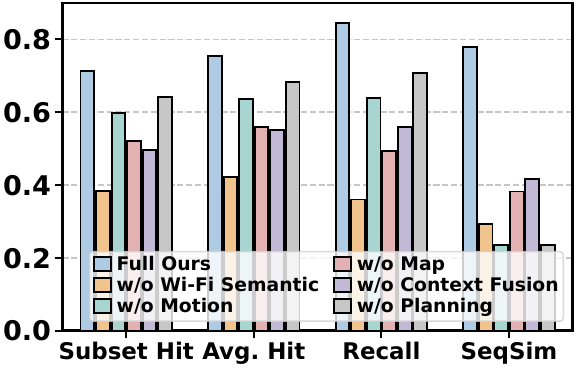}
\vspace{-5mm}
\caption{\textbf{Ablation study.}}\vspace{-5mm}
\label{fig:ablation}
\end{minipage}
\hfill
\begin{minipage}[t]{0.49\linewidth}
\centering
\begin{subfigure}[t]{0.49\linewidth}
\centering
\includegraphics[width=\linewidth]{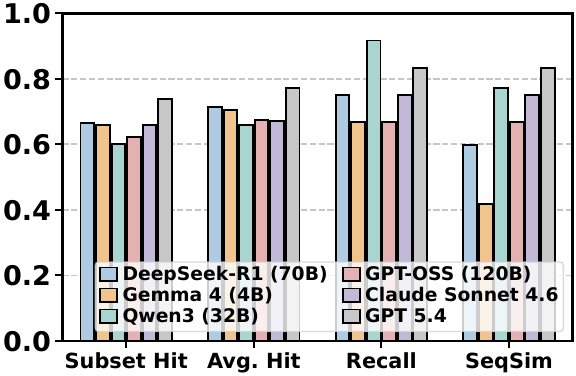}
\vspace{-7mm}
\caption{\textnormal{Trajectory Reconstruction}}\vspace{-5mm}
\label{fig:trajectory-reconstruction}
\end{subfigure}
\hfill
\begin{subfigure}[t]{0.49\linewidth}
\centering
\includegraphics[width=\linewidth]{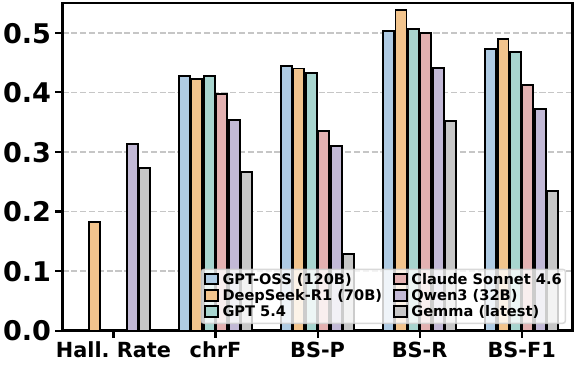}
\vspace{-7mm}
\caption{\textnormal{Diary Generation}}
\label{fig:diary-generation}
\end{subfigure}
\vspace{-4mm}
\caption{\textbf{Cross-model robustness evaluation.}}\vspace{-5mm}
\label{fig:cross-model}
\end{minipage}
\hfill
\begin{minipage}[t]{0.24\linewidth}
\centering
\includegraphics[width=\linewidth]{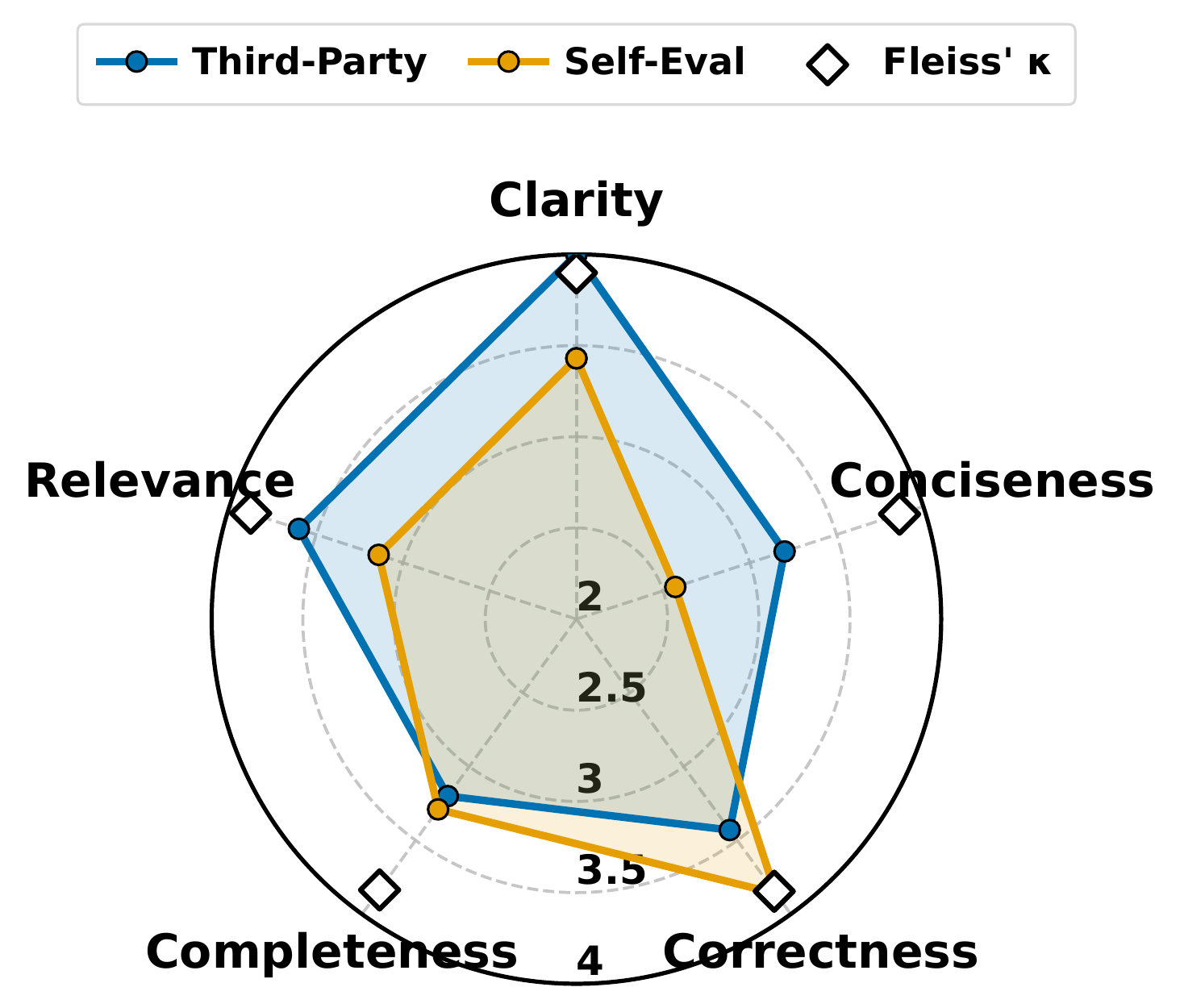}
\vspace{-5mm}
\caption{\textbf{User study results.}}\vspace{-5mm}
\label{fig:user_study}
\end{minipage}

\end{figure*}
\vspace{-1mm}
\subsection{Ablation Study}

Figure~\ref{fig:ablation} shows the contribution of each component to
store-level recovery. Removing the Wi-Fi Semantic Agent causes the largest
degradation, reducing Subset Hit and Avg Hit Ratio from 71.2\%/75.5\% to
38.3\%/42.3\%, and Recall from 84.6\% to 36.0\%. This highlights semantic
Wi-Fi evidence as the key source for identifying visited stores. Removing
the Map Agent or Context Fusion also substantially reduces point-level
recovery, with Avg Hit Ratio dropping to 56.0\% and 55.2\%, respectively.
The Motion Agent mainly improves visit-level consistency, while the
Planning Agent has the smallest effect on point-level metrics. Overall,
the results show that semantic anchoring provides the strongest
contribution, with map, motion, and fusion providing complementary
constraints.

\vspace{-1mm}
\subsection{Cross-Model Evaluation}

We evaluate \sysname with different LLM backbones to assess its robustness to model choice. Figure~\ref{fig:cross-model} shows that trajectory recovery remains relatively stable across backbones, while diary generation is more sensitive to model choice. GPT-5.4 achieves the strongest overall trajectory performance (73.9\% Subset Hit, 77.3\% Avg Hit Ratio, 83.3\% Recall, and 0.833 SeqSim), while Qwen3 (32B) achieves the highest Recall (91.7\%) but a substantially higher hallucination rate (0.313). This gap highlights that high visit recall alone does not guarantee reliable diary generation. Overall, \sysname remains effective across both proprietary and open-weight backbones, demonstrating robustness to the underlying LLM.

\begin{table}[t!]
\centering
\caption{\textbf{Token usage and nominal estimated cost per session.}
\textnormal{Estimates use GPT-4o-equivalent rates of \$2.50/1M input and
\$10.00/1M output tokens; they are not gateway invoices.}}
\label{table:token_usage}
\vspace{-3mm}
\scriptsize

\rowcolors{2}{gray!8}{white}
\begin{tabular}{p{0.32\linewidth}cccc}
\toprule
\textbf{Module} 
& \textbf{Input} 
& \textbf{Output} 
& \textbf{Cost (USD)} 
& \textbf{Calls} \\ 
\midrule

Spatial description
& 22.3k & 1.2k & 6.8$\times10^{-2}$/hr & 1 per floor \\

Wi-Fi semantic agent
& 118.7k & 13.2k & 4.3$\times10^{-1}$/hr & $N_w$ \\

Trajectory validation
& 62.8k & 26.9k & 4.3$\times10^{-1}$/hr & 1 per session \\

Behavior inference
& 1.9k & 0.8k & 1.3$\times10^{-2}$/hr & 1 per session \\

Journal generation
& 3.0k & 1.3k & 2.1$\times10^{-2}$/hr & 1 per session \\

\midrule
\rowcolor{gray!12}
\textbf{Total per session}
& \textbf{208.7k}
& \textbf{43.4k}
& \textbf{9.6$\times10^{-1}$/hr}
& \textbf{All modules} \\

\bottomrule
\end{tabular}
\vspace{-6mm}
\end{table}
\vspace{-1mm}
\subsection{Human Evaluation Study}
\textbf{Setting.} Following AutoLife~\cite{xu2025autolife}, we conduct a user
study to evaluate generated journals along five dimensions: clarity, conciseness,
correctness, completeness, and relevance. Six volunteers rated 11 generated diaries. All dimensions are rated on a 1--4 scale, where 1 indicates that the generated
journal completely does not meet the criterion and 4 indicates that it completely
meets the criterion. Fleiss' $\kappa$ is used to measure inter-rater agreement. For self-evaluation, users assessed the end-to-end system using their own
collected data.

\noindent \textbf{Results.}
Figure~\ref{fig:user_study} shows that the generated journals are generally perceived as clear, relevant, and faithful to the collected data. Evaluations from third-party raters are consistent with the authors' self-assessment, and raters show substantial agreement in their judgments. Correctness emerges as a particular strength, while conciseness remains an area for improvement.
\vspace{-2mm}
\subsection{Token Cost}
\label{subsec:token_cost}
We report nominal API-equivalent cost by logging input and output tokens for
each LLM/VLM call and applying the nominal GPT-4o rates stated in
Table~\ref{table:token_usage}. Requests in our experiments were routed through
a custom third-party API-compatible gateway, and no billing invoice or verified
gateway pricing was available; the reported dollar values are therefore
estimates rather than actual paid cost. Processing one hour of trace consumes
208.7k input tokens and 43.4k output tokens, corresponding to
\$0.96 per hour. In other words, the token cost is roughly \$1 per hour of
processed trajectory. The cost is dominated by Wi-Fi semantic anchoring and
trajectory validation, while behavior inference and journal generation add only
a small fraction of the total.
\vspace{-2mm}
\subsection{Mobile and Edge Deployment}

As location information is privacy sensitive, we evaluate the local deployment cost of \sysname in terms of sensing overhead and on-device processing.

\noindent \textbf{Sensing System Cost.} During daytime collection, the phone only
records passive Wi-Fi scans, IMU samples, timestamps, and optional venue context
using Sensor Logger~\cite{choi2026sensorlogger} on commodity phones (Xiaomi
Flipmix2, vivo S12, and iQoo12). Across the tested phones, a 1-hour trace
consumes 153~mAh (3\% battery) and produces 70~MB of local data on average.
These measurements indicate modest overhead for hour-scale passive sensing.

\noindent \textbf{Running on edge device}. We run our agent with Gemma~4 12B~\cite{google2026gemma4}
locally with LM Studio~\cite{lmstudio2026} on a Mac mini equipped with an M4 chip
and 24~GB memory. Processing a 1-hour daytime trace takes roughly 75~min end-to-end, making the local setup suitable for deferred trajectory analysis and diary generation rather than interactive use. This runtime is therefore suitable for
deferred journal generation rather than interactive use during the trip. The map
can be parsed once offline and reused across visits to the same venue, so map
construction is not included in the per-trace runtime. We further note that
\sysname can also be deployed on mobile devices through the Google AI Edge
Gallery framework~\cite{google2026galleryskills,google2026gemma4edge}; however, phone-side execution
is expected to take longer and is mainly an additional deployment option.

\vspace{-2mm}
\section{Conclusion}
\label{sec:conclusion_discussion}

We have presented the design and implementation of \sysname, an agentic mobile system that converts a floor plan intended for human
interpretation into a machine-readable spatial structure to which ambient Wi-Fi
can be anchored, making store-level indoor logging possible without dedicated
infrastructure. We believe this points to a broader direction: agents that
interpret the spatial artifacts already present in built environments and ground
everyday signals against them.

\begin{acks}
This work was supported in part by the National Science Foundation under award No.~2554332,  No.~2337537, No.~2433914, and No.~2302724.
\end{acks}

\bibliographystyle{ACM-Reference-Format}
\bibliography{main}
\clearpage
\appendix

\section{Map Parsing Evaluation and Failure Analysis}
\label{sec:appendix}
\label{app:map-parsing}

This appendix expands the map-parsing claims made in
Section~\ref{sec:floorplan} and the results reported in
Section~\ref{subsec:map_parsing_evaluation}. The main text gives the headline
comparison; here we explain why the three reported metrics are appropriate,
how to interpret the corresponding scores, and which qualitative cases support
the design choices and limitations discussed in the method.

\subsection{Metric Design and Interpretation}
\label{sec:map-metrics}

Table~\ref{tab:map_parsing_results} in the main text evaluates two questions:
whether AirLog recovers each floor-plan region as a usable polygon, and whether
the recovered polygon receives the correct readable labels. We therefore report
two geometry metrics and one semantic metric.

\noindent\textbf{Area F1: did we recover the regions?}
For an annotated region $g$ and a predicted region $p$, we first compute the two
directed area coverages
\begin{equation}
\begin{aligned}
c_g(g,p) &= \frac{\operatorname{Area}(g\cap p)}
                  {\operatorname{Area}(g)},\\
c_p(g,p) &= \frac{\operatorname{Area}(g\cap p)}
                  {\operatorname{Area}(p)}.
\end{aligned}
\label{eq:area_coverage}
\end{equation}
A pair is eligible when
\begin{equation}
\max(c_g,c_p)\ge 0.5,
\qquad
\min(c_g,c_p)\ge 0.05,
\label{eq:area_eligibility}
\end{equation}
after which we form a label-independent one-to-one matching. Let $TP_A$ be the
matched pairs, $FP_A$ the unmatched predictions, and $FN_A$ the unmatched
annotations. We compute
\begin{equation}
\begin{aligned}
P_A &= \frac{TP_A}{TP_A+FP_A},\\
R_A &= \frac{TP_A}{TP_A+FN_A},\\
\mathrm{AreaF1} &= \frac{2P_AR_A}{P_A+R_A}.
\end{aligned}
\label{eq:area_f1}
\end{equation}

This tolerant criterion reflects the downstream requirement of AirLog: later
modules need one queryable polygon for each floor-plan region, while small
boundary deviations are less important than missing, splitting, or duplicating
a region. Thus, the increase from 0.277 to 0.705 in
Table~\ref{tab:map_parsing_results} indicates substantially more successful
one-to-one region recovery; it should not be interpreted as 70.5\% pixel
overlap.

For reproducibility, this paper-facing Area F1 corresponds to the
\texttt{association\_f1} field in evaluator v8.

\noindent\textbf{Region Purity: did a polygon absorb unrelated content?}
Let $\Omega_{\mathrm{GT}}$ be the union of all evaluator-valid annotated regions.
For each predicted polygon $p$, we compute
\begin{equation}
\mathrm{Purity}(p)=
\frac{\operatorname{Area}(p\cap\Omega_{\mathrm{GT}})}
     {\operatorname{Area}(p)}.
\label{eq:region_purity}
\end{equation}
Predictions receive equal weight within each image, and image-level means are
macro-averaged. Region Purity complements Area F1 because a prediction may
recover the intended region while also absorbing corridor, neighboring-store,
icon, or background pixels. The increase from 0.510 to 0.767 in
Table~\ref{tab:map_parsing_results} therefore indicates cleaner region
predictions with less unrelated content.

\noindent\textbf{Matched Label F1: are the labels correct once geometry is fixed?}
Semantic evaluation first forms a separate label-independent one-to-one
geometry matching at IoU $\ge 0.5$. Label strings are canonicalized identically
for all methods and compared as sets. With pooled semantic true positives,
false positives, and false negatives denoted by $TP_L$, $FP_L$, and $FN_L$, we
report
\begin{equation}
\begin{aligned}
P_L &= \frac{TP_L}{TP_L+FP_L},\\
R_L &= \frac{TP_L}{TP_L+FN_L},\\
\mathrm{LabelF1} &= \frac{2P_LR_L}{P_L+R_L}.
\end{aligned}
\label{eq:label_f1}
\end{equation}

Unmatched polygons are left to the geometry metrics, so geometry failures are
not counted again as semantic errors. Under frozen geometry, the deployed
GPT-4o region-crop configuration obtains 0.814 Matched Label F1, compared with
0.734 for the GPT-5.5 full-map configuration in
Table~\ref{tab:semantic_factorization_main}. Because the backend model and
visual scope change together, this comparison does not isolate the effect of
region cropping or model choice.

\noindent\textbf{Standard-metric sanity check.}
Area F1 uses a task-motivated tolerant matching rule. To verify that the geometry
conclusion does not depend on this rule, we additionally evaluate the same
frozen predictions using standard one-to-one matching at IoU $\geq 0.5$.
Table~\ref{tab:map_standard_metrics} reports Region F1@0.5 and PQ@0.5. The
method ordering remains unchanged under both metrics.

\begin{table}[h!]
\centering
\caption{\textbf{Standard-metric sanity check on map geometry.}
\textnormal{Region F1@0.5 and PQ@0.5 use one-to-one matching at
IoU $\geq 0.5$ and are macro-averaged across the 24 maps.}}
\label{tab:map_standard_metrics}
\vspace{-2mm}
\footnotesize
\setlength{\tabcolsep}{4.2pt}
\begin{tabular}{@{}lcc@{}}
\toprule
\textbf{Method} & \textbf{Region F1@0.5} $\uparrow$ &
\textbf{PQ@0.5} $\uparrow$ \\
\midrule
Color Filling                   & 0.229 & 0.199 \\
Boundary Detection / Su et al. & 0.095 & 0.075 \\
SAM 3                           & 0.020 & 0.016 \\
\textbf{\sysname}              & \textbf{0.590} & \textbf{0.475} \\
\bottomrule
\end{tabular}
\vspace{-2mm}
\end{table}

\subsection{Benchmark and Comparison Protocol}
\label{sec:benchmark_snapshot}

The benchmark used by Section~\ref{subsec:map_parsing_evaluation} contains 24
floor-plan images with 1,488 geometry-valid annotated polygons and 1,400 label
strings; 190 polygons contain multiple readable labels. Twenty images correspond
to publicly discoverable malls with official websites and publicly available
directory or floor-plan images. The remaining four are mall-directory
photographs captured during the same in-the-wild sensing collection used
elsewhere in \sysname. These photographs may contain perspective distortion,
illumination variation, and color shift that are less pronounced in
web-published maps. The benchmark therefore covers both web-published
floor-plan artifacts and photographed mall directories encountered in the
target deployment setting.

The multi-label cases motivate evaluating label sets rather than forcing every
region to contain a single store-name string. Ground-truth polygon geometry was
manually maintained in QGIS against the exact pixel-aligned source image.
GPT-5.5 was used only as an auxiliary tool for label auditing; every final
ground-truth label was manually reviewed against the source floor-plan image.

All reported values use the same archived evaluator-v8 snapshot and prediction
artifacts under OpenCV 4.10.0.  Baseline-specific settings were fixed globally
and applied uniformly across the 24 maps. We did not tune parameters separately
for individual maps or select settings against the benchmark ground truth.

\noindent\textbf{Baseline configurations.}
Color Filling uses gray-world normalization, OKLab clustering with $k=6$, merge
threshold 0.04, random seed 13, and a 30-pixel minimum connected-component area.
PaddleOCR (\texttt{lang=ch}) is applied to the full map, and OCR results are
bound to predicted regions by maximum overlap with center fallback.

Boundary Detection / Su et al. is our implementation of the
edge/region-growing procedure described by Su et al., rather than an official
author-code reproduction. It applies the fixed threshold/edge pipeline at the
original resolution, retains regions of at least 30 pixels, runs OCR on regions
of at least 1,200 pixels among at most the 80 largest candidates, and retries
OCR once after a $90^\circ$ rotation when necessary.

For SAM 3, the configuration reported in
Table~\ref{tab:map_parsing_results} uses the text prompts
\texttt{map region} and \texttt{enclosed spatial region}, longest-side
preprocessing at 2,048 pixels, and a mask-binarization threshold of 0.5.
Separately, the wrapper's omitted object-query score threshold resolves to a
detection-confidence threshold of 0.0 rather than the processor default of
0.5. The configuration uses up to 500 masks, no overlap NMS, and exports each
residual connected component as a polygon.

This permissive configuration produces 24,939 predicted polygons for 1,482
annotated polygons, and 68.05\% of the predicted polygons have no
evaluator-raster overlap with annotated foreground. This large amount of
over-segmentation explains the unusually low 0.037 Area F1 reported for SAM 3.

\noindent\textbf{SAM 3 configuration sensitivity.}
To test whether the low SAM 3 result is primarily an artifact of the permissive
configuration above, we reran all 24 maps using the same checkpoint, text
prompts, resize cap, and OCR pipeline, while restoring the detection-confidence
threshold to 0.5 and retaining one residual component per query. This more
conservative configuration removes the large false-positive fragmentation, but
produces predictions on only one of the 24 maps. With empty-prediction maps
assigned zero, it obtains Region F1@0.5 of 0.042 and PQ@0.5 of 0.033. Thus, the
sensitivity check does not indicate robust SAM 3 performance on this benchmark;
instead, the more conservative configuration suppresses nearly all detections.

Table~\ref{tab:map_parsing_results} is an end-to-end map-parsing comparison:
the visual baselines may differ in both region extraction and label binding.
By contrast, Table~\ref{tab:semantic_factorization_main} freezes the predicted
polygons and visible-label policy, then compares the full-map GPT-5.5 labeling
configuration with the GPT-4o per-region semantic subagent. Because model and
input scope still change together, this table compares the two deployed
configurations but does not independently isolate either factor.

For the GPT-4o per-region run, all 1,920 semantic records were completed from
1,927 application attempts. Seven transient failed attempts affected six
polygons, and all six reached a valid terminal response within the retry budget.
No semantic record was dropped from evaluation.

\subsection{Downstream Map Preparation}
\label{sec:appendix_map_availability}
\label{sec:appendix_map_topology}

\begin{figure}[h!]
    \centering
    \captionsetup[subfigure]{font=footnotesize,skip=2pt}
    \begin{subfigure}[t]{0.48\linewidth}
        \centering
        \includegraphics[width=\linewidth]{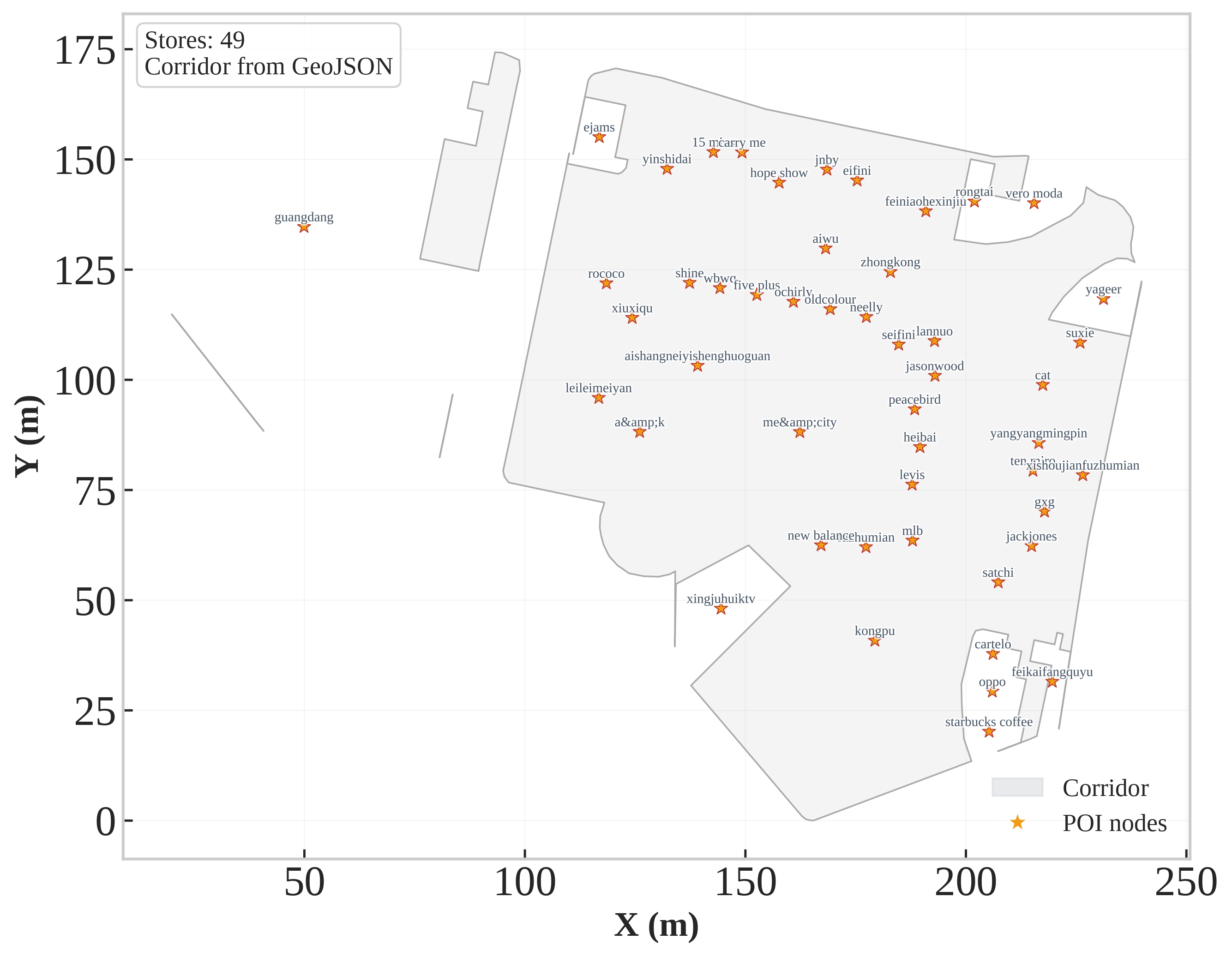}
        \caption{Corridor mask and region targets.}
        \label{fig:map_mask}
    \end{subfigure}\hfill
    \begin{subfigure}[t]{0.48\linewidth}
        \centering
        \includegraphics[width=\linewidth]{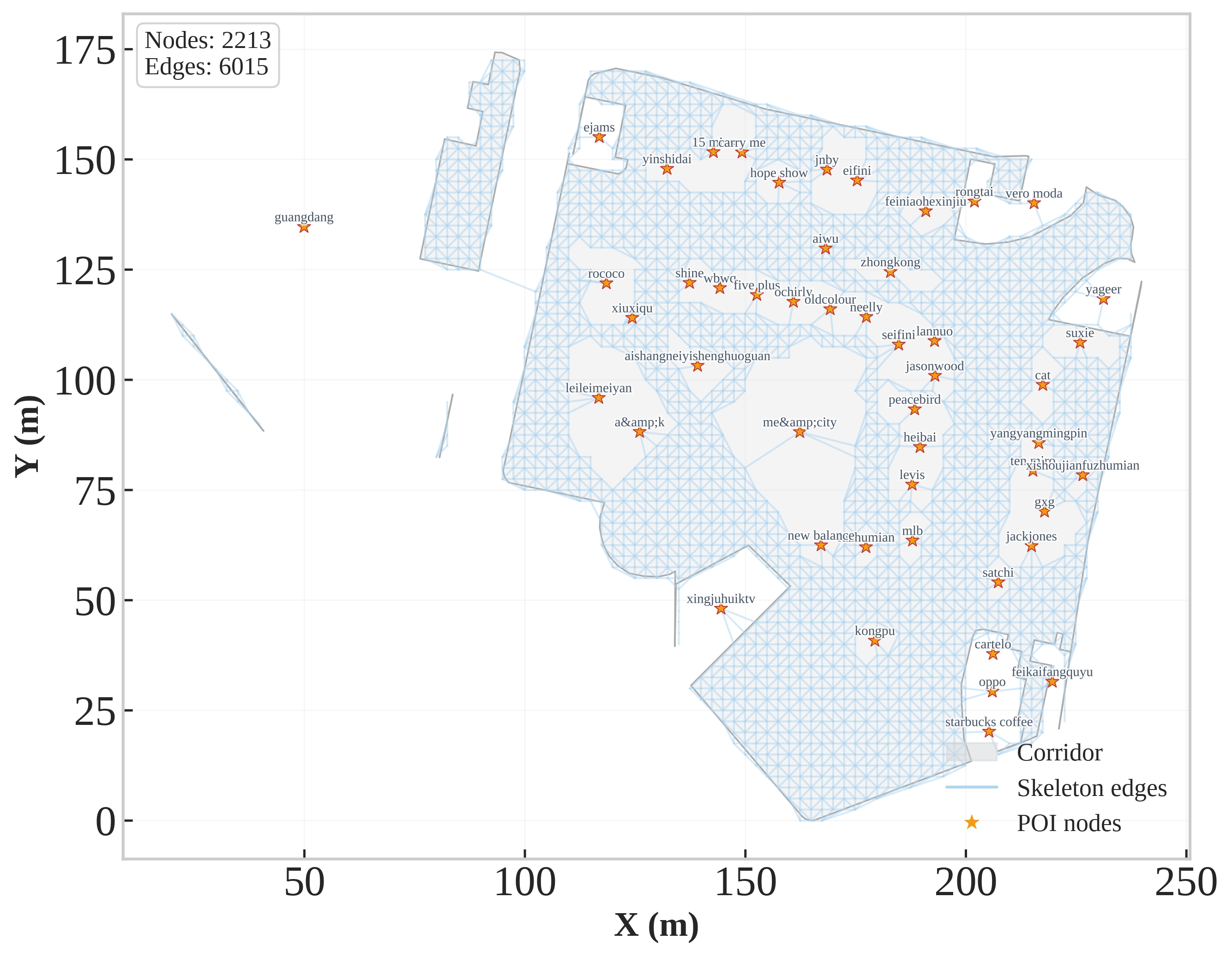}
        \caption{Navigability graph.}
        \label{fig:map_skeleton}
    \end{subfigure}
    \caption{\textbf{Interface between map parsing and trajectory fusion.}
    \textnormal{The labeled GeoJSON region map produced in
    Section~\ref{sec:floorplan} is converted by a separate deterministic stage
    into the corridor mask and navigability graph consumed in
    Section~\ref{sec:spatial_fusion}. These artifacts are not parser outputs and
    are not included in the map-parsing metrics.}}
    \label{fig:overall_map}
    \Description{Two visualizations show a corridor mask with region targets
    and the navigability graph derived later for trajectory fusion.}
\end{figure}

Section~\ref{sec:floorplan} ends with a labeled region map serialized as a
GeoJSON \texttt{FeatureCollection}. Corridor geometry and connectivity are
constructed afterward, before the trajectory stage in
Section~\ref{sec:spatial_fusion}. Figure~\ref{fig:overall_map} visualizes this
interface: the parser-produced regions are converted into a complementary
corridor mask, region targets are placed in the same metric space, and the
corridor is skeletonized into the navigability graph used by the Viterbi
decoder.

This separation also clarifies the scope of
Table~\ref{tab:map_parsing_results}: Area F1, Region Purity, and Matched Label F1
evaluate the parser's region geometry and labels, rather than the downstream
corridor or graph construction.

\FloatBarrier

\subsection{Qualitative Evidence and Remaining Failures}
\label{sec:appendix_map_failures}

Tables~\ref{tab:map_parsing_results} and
\ref{tab:semantic_factorization_main} report the aggregate geometry and label
results. The examples below illustrate specific failure mechanisms discussed in
Section~\ref{sec:floorplan}; they are qualitative evidence rather than
additional quantitative comparisons.

\begin{figure}[!htbp]
    \centering
    \includegraphics[width=\columnwidth]{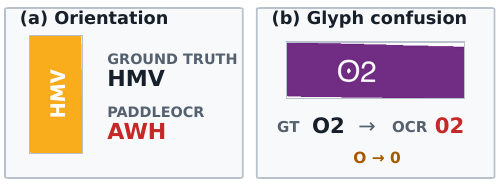}
    \caption{\textbf{Exact-region OCR failures.}
    \textnormal{PaddleOCR reads (a) a visible vertical \textit{HMV} crop as
    \textit{AWH} and (b) \textit{O2} as \textit{02}.}}
    \label{fig:appendix_map_ocr_examples}
    \Description{Two isolated region crops show an HMV-to-AWH orientation
    error and an O2-to-02 glyph-confusion error.}
\end{figure}

\noindent\textbf{Exact-region OCR errors.}
Figure~\ref{fig:appendix_map_ocr_examples} fixes the target geometry and gives
PaddleOCR only the exact region crop. The readable vertical \textit{HMV} label
is returned as \textit{AWH}, while \textit{O2} is returned as \textit{02}.
Region isolation therefore removes spatial-assignment ambiguity but does not,
by itself, resolve OCR orientation or glyph confusion.

\begin{figure}[!htbp]
    \centering
    \includegraphics[width=\columnwidth]{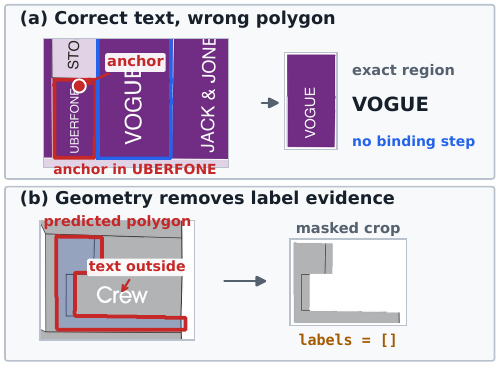}
    \caption{\textbf{Two map-parsing interface failures.}
    \textnormal{(a) A cached full-map GPT-5.5 result recognizes
    \textit{VOGUE} but places its coordinate in the adjacent
    \textit{UBERFONE} polygon; an exact-region input removes this binding step.
    (b) For one \textit{Crew}-associated component, the predicted polygon
    removes the readable word from the masked crop, leaving no readable
    evidence for the semantic subagent.}}
    \label{fig:appendix_map_mechanisms}
    \Description{The upper panel contrasts a misplaced VOGUE coordinate with
    an exact-region crop. The lower panel shows a predicted polygon excluding
    Crew, followed by a text-free masked crop and an empty semantic result.}
\end{figure}

\noindent\textbf{Full-map coordinate-binding error.}
Figure~\ref{fig:appendix_map_mechanisms}(a) provides an illustrative cached
full-map failure case. GPT-5.5 recognizes \textit{VOGUE}, yet its predicted
coordinate falls inside the adjacent \textit{UBERFONE} polygon. Per-region
labeling removes this additional coordinate-prediction and polygon-binding
step. This example illustrates the failure mechanism and is not used for the
quantitative comparison.

\noindent\textbf{Geometry-to-semantics cascade.}
Figure~\ref{fig:appendix_map_mechanisms}(b) illustrates a remaining limitation
of the factorized pipeline. For one matched \textit{Crew}-associated component,
the predicted polygon excludes the readable word. The resulting polygon-masked
crop therefore removes the visual evidence needed by the semantic subagent.
The adjacent \textit{Crew} component, whose polygon retains the text, is labeled
correctly.

\FloatBarrier
\section{Wi-Fi Semantic Evidence and Collection Context}
\label{sec:appendix_wifi_evidence}
This appendix groups the evidence supporting the use of ambient Wi-Fi as a weak
semantic cue: collection coverage, the sparsity and failure modes of lexical
SSID--POI matching, and the ambiguity of RSSI-based nearest-signal decisions.

\subsection{Collection Coverage}
\label{sec:appendix_collection_context}

\begin{table}[t]
\centering
\caption{\textbf{Dataset statistics across six cities.}}
\label{tab:dataset_stats}
\vspace{-2mm}

\rowcolors{2}{gray!10}{white}
\resizebox{\columnwidth}{!}{
\begin{tabular}{lrrrrrrr}
\hline
\textbf{Metric} & \textbf{Hong Kong} & \textbf{Beijing} & \textbf{Shanghai} &
\textbf{Shenzhen} & \textbf{Dalian} & \textbf{Nanjing} & \textbf{Total} \\
\hline

Active hours (h)      & 59.3    & 5.8    & 208.4   & 225.5   & 2.6    & 101.3   & \textbf{603.0} \\
GPS points            & 157,360 & 2,137  & 277,356 & 319,865 & 6,696  & 168,041 & 931,455 \\
Wi-Fi observations    & 260,394 & 2,711  & 25,798  & 45,801  & 10,391 & 29,472  & 374,567 \\
Unique SSIDs (raw)    & 21,303  & 984    & 1,730   & 2,125   & 1,604  & 1,274   & 29,020 \\
Unique BSSIDs (raw)   & 56,615  & 1,742  & 2,541   & 3,199   & 2,832  & 1,922   & 68,851 \\
Avg.\ SSIDs per scan  & 81.0    & 46.7   & 3.0     & 5.9     & 48.8   & 6.2     & 59.4 \\
\hline
\end{tabular}}
\end{table}

\begin{figure*}[t]
\centering
\captionsetup[subfigure]{font=footnotesize,skip=2pt}
\begin{subfigure}[t]{0.29\textwidth}
    \centering
    \includegraphics[width=\linewidth,height=3.7cm,keepaspectratio]{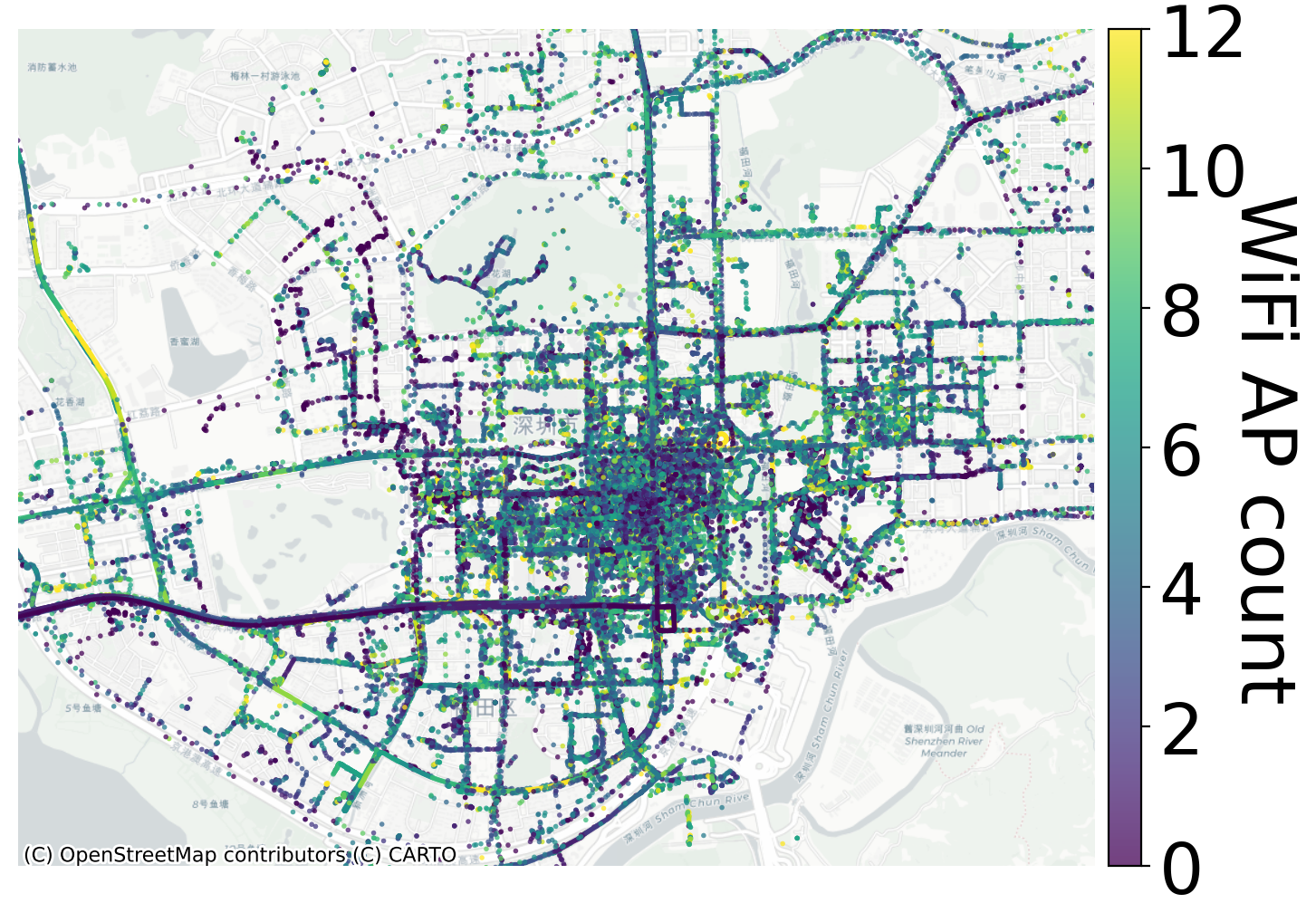}
    \caption{Shenzhen density.}
    \label{fig:wifi_shenzhen}
\end{subfigure}\hfill
\begin{subfigure}[t]{0.29\textwidth}
    \centering
    \includegraphics[width=\linewidth,height=3.7cm,keepaspectratio]{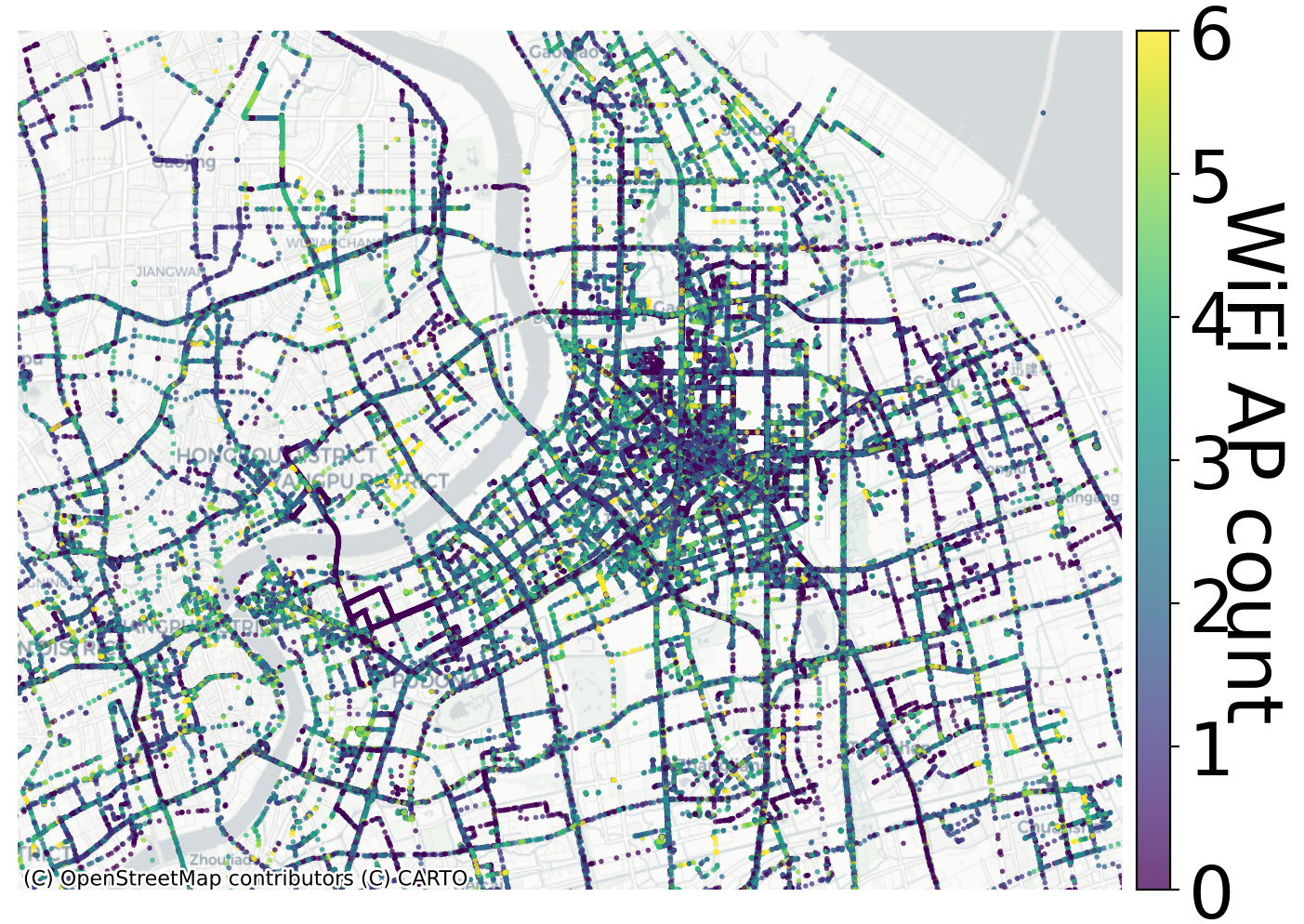}
    \caption{Shanghai density.}
    \label{fig:wifi_shanghai}
\end{subfigure}\hfill
\begin{subfigure}[t]{0.38\textwidth}
    \centering
    \includegraphics[width=\linewidth,height=3.7cm,keepaspectratio]{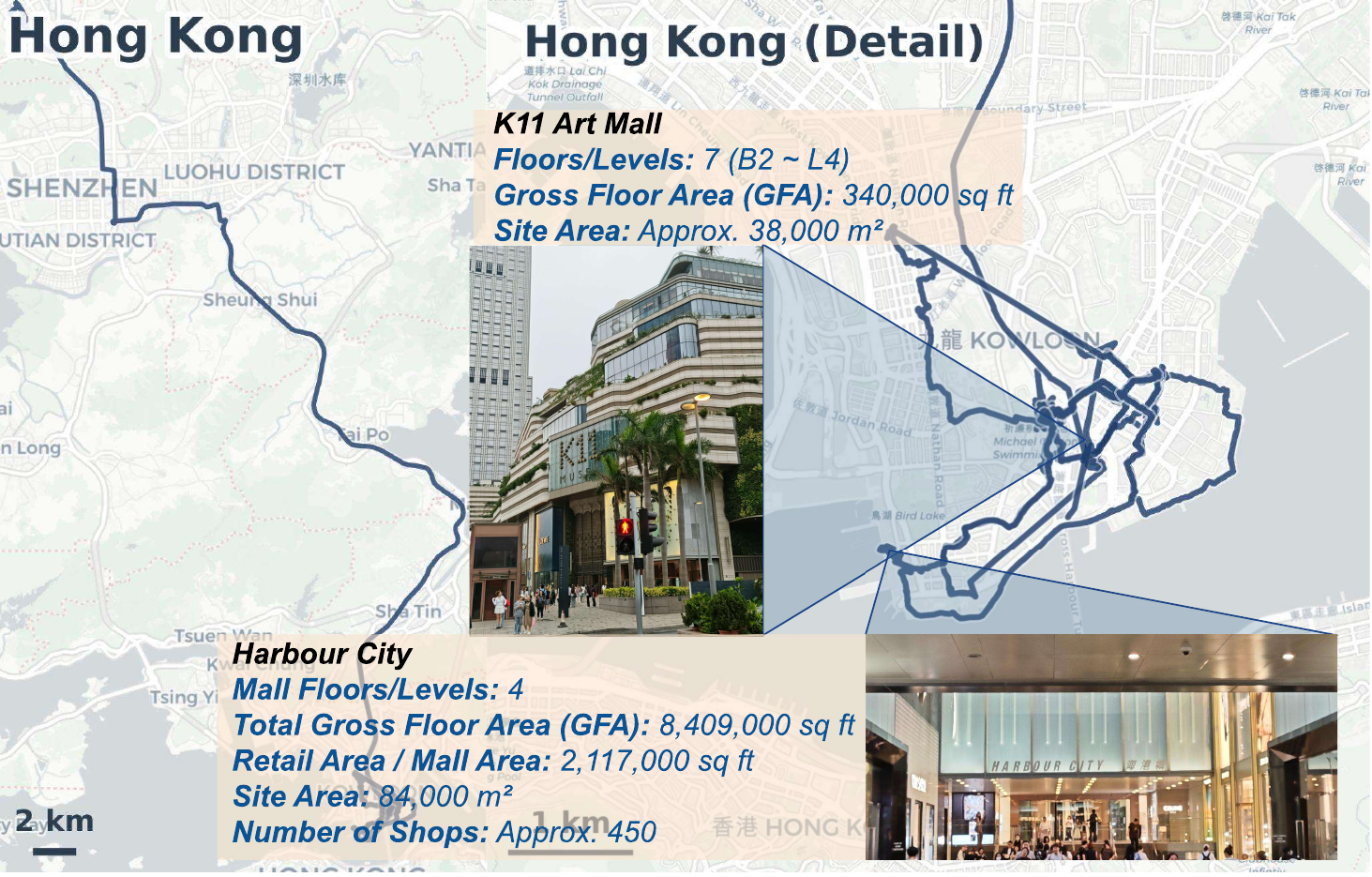}
    \caption{Hong Kong session context.}
    \label{fig:hk_trajectories}
\end{subfigure}
\caption{\textbf{Representative Wi-Fi collection context.}
\textnormal{The density maps summarize two urban collection areas, while the
Hong Kong panel provides a session-level view of mobility coverage.}}
\label{fig:wifi_collection_context}
\label{fig:wifi_density}
\end{figure*}

Figure~\ref{fig:wifi_collection_context} summarizes representative urban and
session-level coverage. These figures document collection context; store-level
reconstruction remains grounded in indoor Wi-Fi observations, PDR motion
continuity, and structured indoor topology.

\subsection{Lexical Similarity and Representative Failure Cases}
\label{sec:appendix_wifi_lexical}
\begin{table}[t]
\centering
\caption{\textbf{Number of SSID--POI pairs by similarity range and precision.}
\textnormal{Precision is computed over manually annotated same-store labels.}}
\label{tab:ssid_similarity_distribution}
\renewcommand{\arraystretch}{1.12}\footnotesize
\begin{tabular}{lccc}
\toprule
\textbf{Range} & \textbf{Cosine (Char)} & \textbf{Cosine (Word)} & \textbf{Cosine (TF-IDF)} \\
\midrule
0.0      & 9,731   & 184,841 & 184,871 \\
0.0--0.2 & 33,454  & 3       & 3       \\
0.2--0.4 & 72,318  & 131     & 65      \\
0.4--0.6 & 54,474  & 137~(34.3\%) & 100~(9.0\%) \\
0.6--0.8 & 14,516  & 17~(58.8\%) & 61~(44.3\%) \\
0.8--1.0 & 647~(52.2\%) & 11~(100\%) & 40~(90.0\%) \\
\midrule
\textbf{Total} & \textbf{185,140} & \textbf{185,140} & \textbf{185,140} \\
\bottomrule
\end{tabular}
\end{table}

The character-level distribution shows a clear non-zero tail, indicating that
many SSIDs still contain partial POI-related strings. However, only a small
fraction of pairs reach strong lexical overlap: 647 pairs exceed 0.8 at
character level with a precision of 52.2\%, and 14,516 additional pairs fall
between 0.6 and 0.8. Among word-level and TF-IDF pairs, high-similarity matches
($>$0.8) achieve up to 100\% and 90.0\% precision respectively, but account for
only 11 and 40 pairs out of 185,140, showing that exact token matches are
extremely sparse in practice.

Table~\ref{tab:bad_cases} illustrates a complementary failure mode: surface
string similarity can rank misleading SSID--POI pairs highly even when the
semantic explanation points elsewhere. These examples justify using semantic
reasoning over names, local context, and map constraints; they are not presented
as a standalone quantitative benchmark.

\begin{table}[t]
\centering
\caption{\textbf{Representative bad cases of Wi-Fi SSID--POI matching.}
\textnormal{String similarity algorithms fail where LLM semantic reasoning
succeeds.}}
\label{tab:bad_cases}
\footnotesize
\renewcommand{\arraystretch}{1.12}
\begin{tabular}{llc}
\toprule
\textbf{Wi-Fi SSID} & \textbf{POI} & \textbf{Best Sim.} \\
\midrule
SKH-SchoolDevice & SKH Holy Carpenter Church & 0.693 \\
Beauty           & A Boutea                  & 0.833 \\
MNSGUEST         & Monster Sushi             & 0.820 \\
\bottomrule
\multicolumn{3}{p{7.5cm}}{\footnotesize $^a$~\textit{SKH-SchoolDevice}: SKH is a dominant institutional prefix of the Hong Kong Sheng Kung Hui; ``SchoolDevice'' identifies a campus device network, pointing to an affiliated lodging facility rather than the church itself.} \\
\multicolumn{3}{p{7.5cm}}{\footnotesize $^b$~\textit{Beauty}: ``A Boutea'' is a beverage brand whose name superficially resembles ``Beauty'' in spelling; the SSID more plausibly belongs to a cosmetics or personal-care venue.} \\
\multicolumn{3}{p{7.5cm}}{\footnotesize $^c$~\textit{MNSGUEST}: MNS follows a typical three-letter property code convention; combined with its location in the Miramar cluster in Tsim Sha Tsui, the network is better attributed to Miramar Shopping Centre's guest Wi-Fi than to a restaurant abbreviation.} \\
\end{tabular}
\end{table}

\subsection{RSSI Ambiguity}
\label{sec:appendix_wifi_ambiguity}
\begin{figure}[t]
    \centering
    \captionsetup[subfigure]{font=footnotesize,skip=2pt}
    \begin{subfigure}[t]{0.4\textwidth}
        \centering\includegraphics[width=\linewidth]{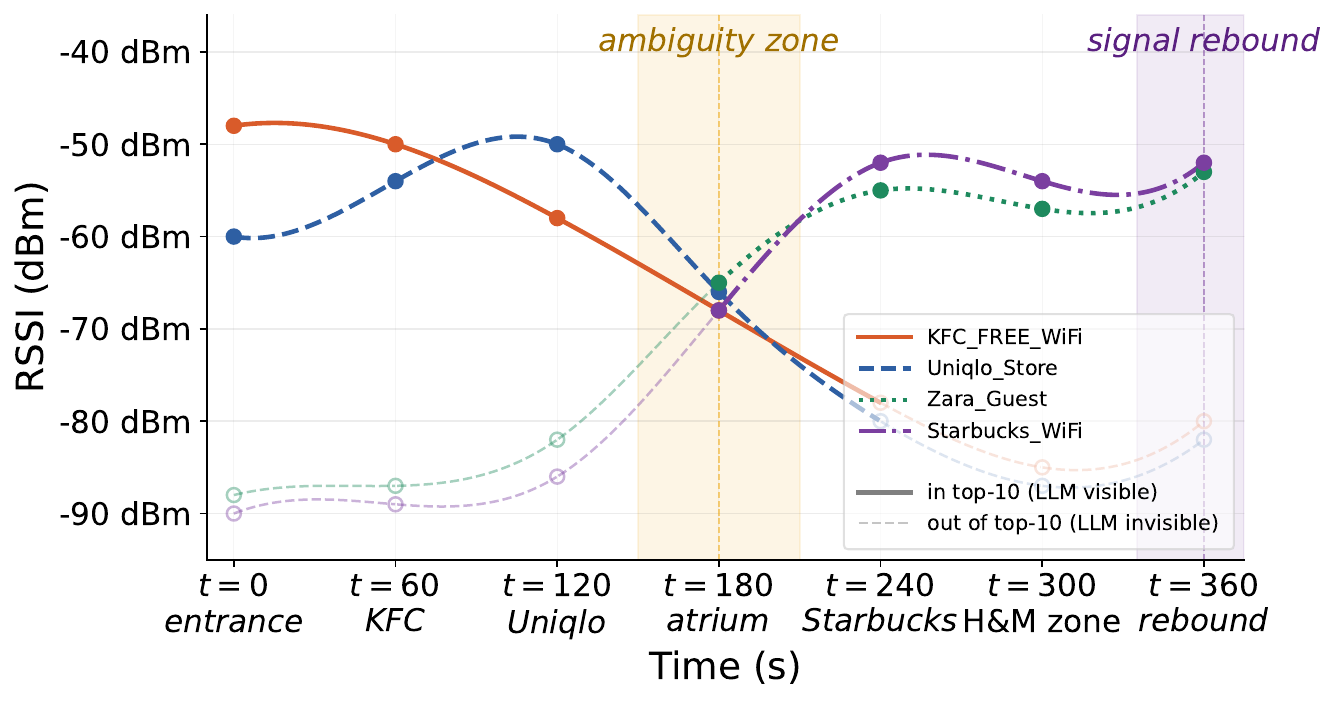}
        \caption{\textbf{RSSI variation.} Solid lines indicate SSIDs in the
        top-$K$ window; faded dashed lines indicate lower-ranked signals.}
        \label{fig:rssi_trajectory}
    \end{subfigure}\hfill
    \begin{subfigure}[t]{0.35\textwidth}
\centering\includegraphics[width=\linewidth,height=5.0cm,keepaspectratio]{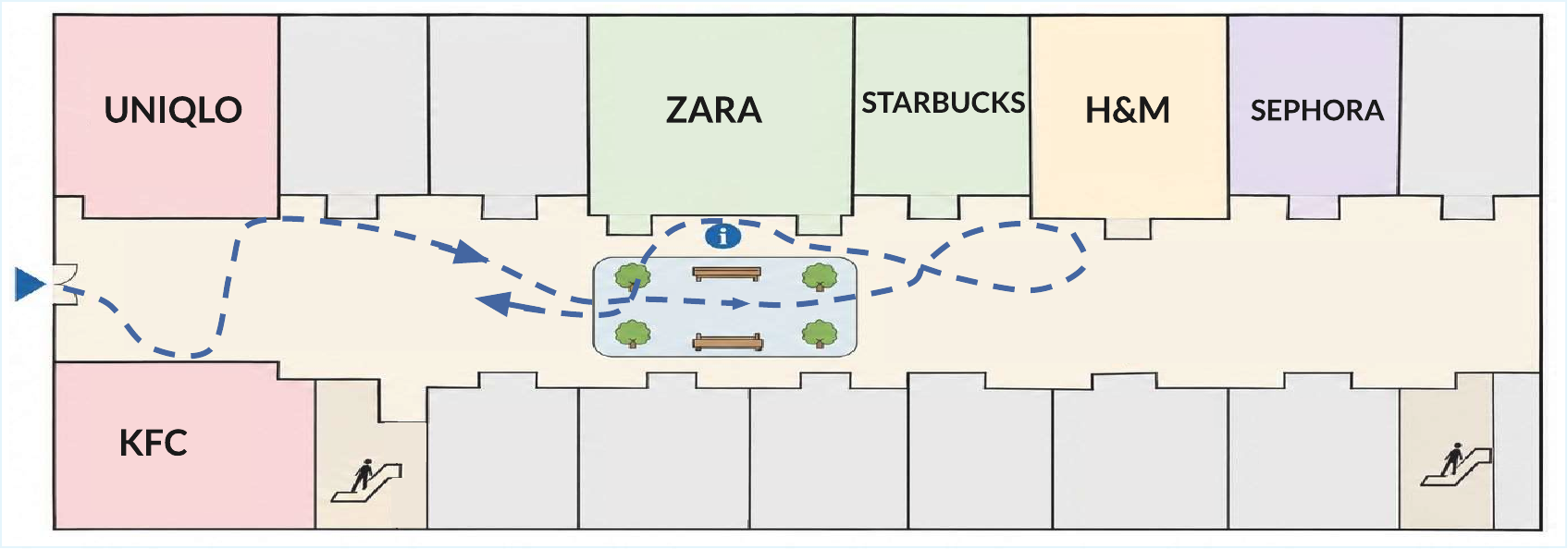}
        \caption{\textbf{Spatial layout.} The corresponding mall trajectory
        remains compatible with multiple nearby stores.}
        \label{fig:semantic_anchor_map}
    \end{subfigure}
    \caption{\textbf{RSSI ambiguity and spatial layout.}
    \textnormal{The profiles and map show why semantic anchoring is formulated
    as soft multi-label inference rather than a deterministic nearest-signal
    decision.}}
    \label{fig:semantic_anchor_full}
\end{figure}

Figure~\ref{fig:semantic_anchor_full} expands the Wi-Fi anchoring example from
the main paper. Nearby access points overlap substantially along a single walk,
and one signal window can remain compatible with multiple neighboring stores.
This motivates treating the top-$K$ SSIDs as soft evidence rather than selecting
the strongest signal as the user's location.

\section{Additional Journal Examples}
\label{sec:appendix_diary_examples}
The following examples supplement the journal comparison in the main paper.
\underline{\textit{Underlined italic}} denotes ground-truth POI names, and
{\color{red} red} denotes hallucinated or incorrect content.

\begin{figure}[H]
    \centering
    \includegraphics[width=\linewidth]{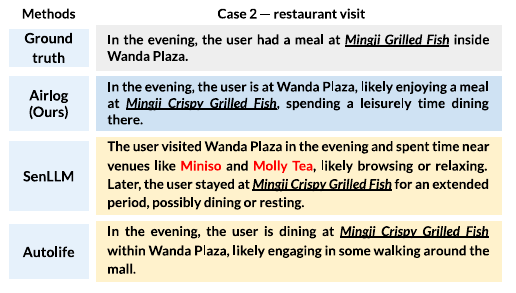}
    \caption{\textbf{Journal generation case 2: restaurant visit.}}
    \label{fig:diary_case_2}
\end{figure}

\begin{figure}[H]
    \centering
    \includegraphics[width=\linewidth]{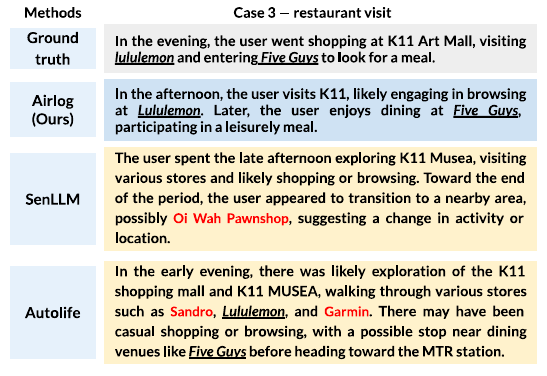}
    \caption{\textbf{Journal generation case 3: shopping and dining visit.}}
    \label{fig:diary_case_3}
\end{figure}

\clearpage
\section{Prompt Templates}
\label{sec:appendix_prompt}

\subsection{Spatial Description Generation}
\label{app:spatial_description_prompt}
This prompt converts GeoJSON store features into normalized names, neighbor
relations, and natural-language position descriptions. The output is a compact
spatial context that later prompts can use without directly reasoning over raw
polygon coordinates.

\begin{figure}[H]
    \centering
    \includegraphics[width=\linewidth,height=0.66\textheight,keepaspectratio]{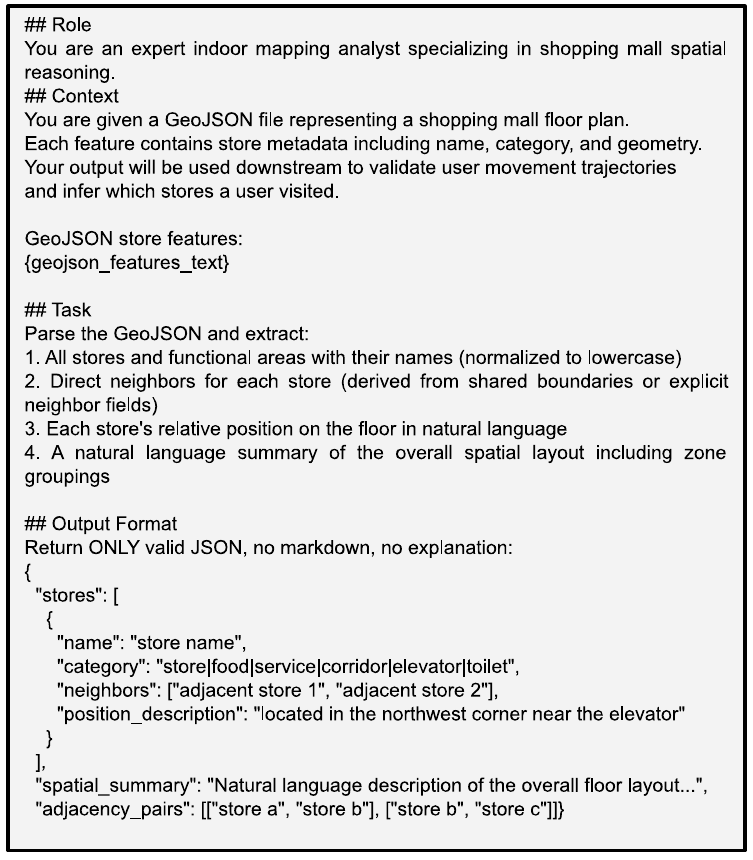}
    \caption{\textbf{Prompt template for spatial description generation from
    GeoJSON floor-plan features.}}
    \label{fig:spatial_description_prompt}
\end{figure}

\newpage
\subsection{Wi-Fi Semantic Anchoring}
\label{app:wifi_semantic_prompt}
The following prompt maps Wi-Fi SSIDs to candidate store locations via semantic
similarity, relative signal strength, and spatial context.

\begin{figure}[H]
    \centering
    \includegraphics[width=\linewidth,height=0.66\textheight,keepaspectratio]{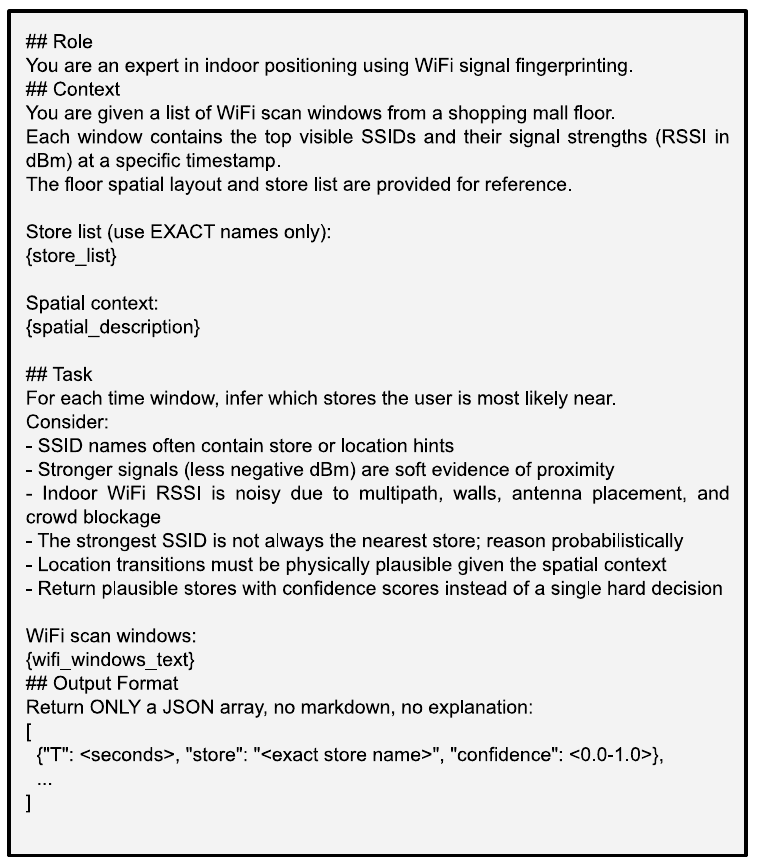}
    \caption{\textbf{Prompt template for Wi-Fi semantic anchoring using SSID
    semantics and signal evidence.}}
    \label{fig:wifi_semantic_prompt}
\end{figure}

\end{document}